\documentclass[10pt,twocolumn,letterpaper]{article}

\usepackage[pagenumbers]{cvpr} 

\usepackage{microtype}

\renewcommand{\paragraph}[1]{\vspace{.5em}\noindent\textbf{#1.}}

\usepackage{soul}
\setuldepth{foobar}

\usepackage{tikz}
\usepackage{amsmath}
\usepackage{array}
\usepackage{stfloats}
\usepackage{url}

\usepackage{amsthm}
\usepackage{booktabs}
\usepackage{amssymb}

\usepackage{multirow}
\usepackage{flexisym}
\usepackage{breqn}
\usepackage{amsfonts}
\usepackage[pdftex]{thumbpdf}
\usepackage{newfloat}
\usepackage{pgffor}
\usepackage{epsfig,endnotes}

\usepackage{lscape}

\usepackage{pgfplots, pgfplotstable}
\pgfplotsset{width=7.5cm,compat=1.16}
\pgfplotsset{every y tick label/.append style={font=\tiny, xshift=0.5ex}}
\pgfplotsset{every x tick label/.append style={font=\tiny}}

\graphicspath{{images/}{pdfs/}}
\DeclareGraphicsExtensions{.pdf,.jpeg,.png}

\newtheorem{definition}{Definition}

\definecolor{cvprblue}{rgb}{0.21,0.49,0.74}
\usepackage{hyperref}
\hypersetup{pagebackref,breaklinks,colorlinks,allcolors=cvprblue}

\def\paperID{32529} 
\def\confName{CVPR}
\def\confYear{2026}

\title{XFED: Non-Collusive Model Poisoning Attack Against Byzantine-Robust Federated Classifiers}

\author{
\begin{tabular}{c}
Israt Jahan Mouri\textsuperscript{1}, Muhammad Ridowan\textsuperscript{2}, Muhammad Abdullah Adnan\textsuperscript{1}\\
\textsuperscript{1}Bangladesh University of Engineering and Technology (BUET), Dhaka, Bangladesh\\
\textsuperscript{2}TigerIT Bangladesh Ltd., Dhaka, Bangladesh\\
{\tt\small 0421054001@grad.cse.buet.ac.bd, 
ridowan007@gmail.com,
adnan@cse.buet.ac.bd}
\end{tabular}
}

\begin{document}
\maketitle
\begin{abstract}
Model poisoning attacks pose a significant security threat to Federated Learning (FL). Most existing model poisoning attacks rely on collusion, requiring adversarial clients to coordinate by exchanging local benign models and synchronizing the generation of their poisoned updates. However, sustaining such coordination is increasingly impractical in real-world FL deployments, as it effectively requires botnet-like control over many devices. This approach is costly to maintain and highly vulnerable to detection. This context raises a fundamental question: Can model poisoning attacks remain effective without any communication between attackers? To address this challenge, we introduce and formalize the \textbf{non-collusive attack model}, in which all compromised clients share a common adversarial objective but operate independently. Under this model, each attacker generates its malicious update without communicating with other adversaries, accessing other clients’ updates, or relying on any knowledge of server-side defenses. To demonstrate the feasibility of this threat model, we propose \textbf{XFED}, the first aggregation-agnostic, non-collusive model poisoning attack. Our empirical evaluation across six benchmark datasets shows that XFED bypasses eight state-of-the-art defenses and outperforms six existing model poisoning attacks. These findings indicate that FL systems are substantially less secure than previously believed and underscore the urgent need for more robust and practical defense mechanisms.
\end{abstract}

\section{Introduction}

Federated Learning (FL)~\cite{konevcny2016federated,konevcny2016federatedStrategies} enables multiple clients to train a shared model collaboratively without exchanging their private data. In this framework, each client computes an update using its own data and sends it to a central server. The server then aggregates these updates to create a global model, which is redistributed to all clients. While it is a promising solution, FL systems are vulnerable to model poisoning attacks due to their distributed nature \cite{shejwalkar2021manipulating,bhagoji2019analyzing,fang2020local,baruch2019little,bulyan,xie2020fall,bagdasaryan2020backdoor}. Model poisoning attacks involve direct manipulation of the local model gradients on malicious devices before sharing them with the central server in each training epoch.

To address the vulnerabilities posed by poisoning attacks, the machine learning community has introduced several Byzantine-robust aggregation rules~\cite{krum,trimmed_mean_median, shejwalkar2021manipulating, li2023experimental, xu2022byzantine, bulyan, fang2020local, cao2021fltrust}. These rules are specifically designed to mitigate the impact of adversarial clients by identifying and excluding malicious updates~\cite{trimmed_mean_median, bernstein2018signsgd, krum, bulyan, shejwalkar2021manipulating, li2019rsa, sun2019can, hongyan2019cronus, lin2020ensemble, diakonikolas2019sever, diakonikolas2017being, chen2018draco, konstantinidis2021byzshield, rajput2019detox}. These aggregation rules ensure that only benign updates contribute to the global model. However, despite the robustness of Byzantine-robust aggregation rules, recent studies \cite{fang2020local, shejwalkar2021manipulating, shejwalkar2022back, baruch2019little} have shown that carefully crafted adversarial model updates can circumvent them. These Byzantine-robust attacks assume that attackers collude by sharing statistical insights from their local models to create unified malicious updates to amplify the impact of their attacks. This coordinated effort makes the poisonous updates closely resemble the benign distribution, enabling them to bypass Byzantine-robust aggregation filters.

As highlighted in previous research, successfully launching a poisoning attack in real-world FL systems typically requires compromising around 20\% of the total clients \cite{shejwalkar2021manipulating, fang2020local}. In an FL system with 1 million participants, this means an attacker would need control over approximately 200,000 devices. However, beyond achieving this large-scale compromise, the attacker must also ensure reliable communication among the compromised devices to generate consistent malicious updates, often using a botnet or a command-and-control (C\&C) server. Maintaining a botnet of this scale presents significant logistical complexities, reliability issues, and ongoing maintenance challenges. The considerable effort and resources needed to sustain such a botnet make collusion highly impractical. Furthermore, communication among a large number of devices introduces additional risks, including detection by firewalls and the difficulty of maintaining synchronization among devices. Given these challenges, Shejwalkar et al. \cite{shejwalkar2022back} argue that collusive attacks are impractical due to the high costs of coordination among attackers and the risks of detection. To address these concerns, recent studies, such as MPAF \cite{cao2022mpaf} and PoisonedFL \cite{xie2024model}, propose creating attacker-controlled fake clients that mimic real participants, allowing the adversary to reach the same nominal fraction of attackers ($\approx20\%$) without compromising an impractically large number of physical devices. However, these fake clients face their own issues, as they must avoid detection by client verification mechanisms. This requirement can limit their viability in secure or verification-sensitive federated learning deployments. In this paper, we introduce an alternative strategy: a \textit{silent non-collusive attack} in which compromised devices act independently and require no inter-attacker communication.

\paragraph{Non-collusive attack} We propose a practical adversarial model in which attackers operate independently, without explicit coordination in each round. An attacker can release malware, exploit widespread application vulnerabilities, or use compromised software updates to carry out an attack campaign, potentially compromising a significant number of clients. There are real-world instances of such silent malware attacks \cite{sharma2025characterization,langner2011stuxnet,shi2020vahunt}; therefore, we believe this approach is more realistic in achieving a significant proportion of clients than using fake clients. Each compromised client independently creates its own malicious model updates, without communicating with other adversarial clients. To the best of our knowledge, this is the first work to investigate non-collusive model poisoning attacks. However, crafting a non-collusive model poisoning attack poses significant challenges compared to collusive attacks. A non-collusive attacker has limited knowledge, restricted to the information available on their compromised device. The lack of access to a broader distribution of benign or adversarial updates makes it particularly difficult for a non-collusive adversary to mount an effective Byzantine-robust attack.

\paragraph{XFED: A Practical Non-Collusive Model Poisoning Attack} As a proof-of-concept for non-collusive attacks, we introduce XFED, a novel model poisoning attack that operates under a more realistic threat model. XFED is designed to be \textbf{\textit{aggregation-agnostic}} \cite{baruch2019little, shejwalkar2021manipulating}, it operates without any knowledge of the server’s aggregation rule. The high-level procedure of XFED is as follows: the adversary first computes a benign model update using local data from the compromised device. Next, the attacker selects a predefined malicious perturbation vector to define a malicious direction in the gradient space. The malicious update is then generated by applying this perturbation vector, scaled by a scaling coefficient, to the benign update. Determining an appropriate scaling coefficient is critical to the success of the attack. To maximize impact while avoiding detection, the scaling must be large enough to influence the global model in the attacker’s intended direction but subtle enough to remain within the statistical range of benign updates. The attacker estimates the scale using only the sequence of global models observed up to the current training round.

\paragraph{Evaluation} We evaluate our XFED framework using six benchmark datasets. Our results show that, under the least knowledge scenario, where the attacker has no information about the deployed defense, nor access to the local data or models of either compromised or benign clients, XFED successfully compromises five Byzantine-robust aggregation rules and three state-of-the-art FL defenses~\cite{krum, trimmed_mean_median, li2023experimental, xu2022byzantine, nguyen2022flame, fung2018mitigating, fereidoonifreqfed}) in different FL settings. We further demonstrate that, although the use of a trusted root dataset can help the server mitigate model poisoning attacks \cite{cao2021fltrust}, this approach violates a fundamental design principle of FL, which prohibits the server from storing any client data. These findings highlight the urgent need for more robust and practical defense mechanisms. All experiments are fully reproducible, and the implementation is open-source, developed in Python and hosted on Google Colab\footnote{\url{https://github.com/pkse-searcher/xfed}}.

\section{Background and Related Work}\label{background}

\subsection{\textbf{Federated Learning System}}

We consider a standard Federated Learning setup~\cite{konevcny2016federated,konevcny2016federatedStrategies,mcmahan2017communication}, where \( k \) clients collaboratively train a model. The server initializes the global parameters \( \boldsymbol{\theta}^{(0)} \) and, at each round \( t \), broadcasts \( \boldsymbol{\theta}^{(t)} \) to all clients. Each client \( i \) updates the model using its local dataset \( \mathcal{D}_i \) and computes \( ^{(t+1)}\boldsymbol{\phi}_i = \boldsymbol{\theta}^{(t)} - \eta \nabla \mathcal{L}_i(\boldsymbol{\theta}^{(t)}; \mathcal{D}_i) \), 
where \( \eta \) is the learning rate. Clients send their updates \( ^{(t+1)}\boldsymbol{\phi}_i\) to the server, which aggregates them to produce the next global model: \( \boldsymbol{\theta}^{(t+1)} = \mathrm{aggregation}(^{(t+1)}\boldsymbol{\phi}_1, \ldots, ^{(t+1)}\boldsymbol{\phi}_k) \). This process continues for \( T \) rounds until convergence. Federated Learning commonly operates in two settings: (1) \textbf{\textit{Cross-device}}~\cite{rehman-cross-device,cross-device}, which involves a huge number of clients (often millions), of which only a small subset participates in each round; and (2) \textbf{\textit{Cross-silo}}~\cite{huang2022crosssilo,huangCross-silo}, which involves a moderate number of clients (typically tens to hundreds), all of whom participate in every round.

\subsection{\textbf{Poisoning Attacks on FL}}
FL is known to be vulnerable to various poisoning attacks due to its distributed nature \cite{blanchard2017machine, baruch2019little, bhagoji2019analyzing, bagdasaryan2020backdoor, fang2020local, bulyan, mahloujifar2019universal, xie2020fall, munoz2017towards, jagielski2018manipulating}. Poisoning attacks pose a significant risk to federated systems \cite{munoz2017towards}, as they involve attackers sending harmful model updates to the central server. Attacks in FL can be categorized according to the objectives of the adversary into two main types: (1) \textit{\textbf{untargeted}} and (2) \textit{\textbf{targeted}} attacks. In untargeted poisoning attacks, the goal is to maximize misclassification across all test inputs, thereby significantly reducing the global model's accuracy~\cite{shejwalkar2021manipulating, fang2020local, baruch2019little, bulyan, mahloujifar2019universal, xie2020fall}. On the other hand, targeted poisoning attacks aim to achieve specific classification results for certain data while maintaining high accuracy on other inputs ~\cite{wang2020attack, bagdasaryan2020backdoor, bhagoji2019analyzing, tolpegin2020data}. While targeted attacks can compromise particular aspects of the global model, untargeted attacks pose a more severe threat by potentially crippling the model entirely. Poisoning attacks in FL are further classified into two types based on the capabilities of the adversary: (1) \textit{\textbf{data poisoning attacks}} and (2) \textit{\textbf{model poisoning attacks}}. In data poisoning attacks, the adversary modifies the training data on compromised devices and indirectly influences the model~\cite{jagielski2018manipulating, tolpegin2020data,sun2021data}. On the other hand, model poisoning attacks involve altering the local model on compromised devices before sharing them with the central server during training~\cite{shejwalkar2021manipulating,bhagoji2019analyzing,fang2020local,baruch2019little,bulyan,xie2020fall,bagdasaryan2020backdoor}. However, data poisoning attacks have proven to be only minimally effective in reducing the global model's accuracy~\cite{mouri23towards, mouri2024data,fang2020local, shejwalkar2022back}. In this paper, we focus on the model poisoning attacks.


\paragraph{Byzantine Robust Aggregations} To address the vulnerabilities posed by poisoning attacks, the machine learning community has introduced several Byzantine-robust aggregation rules~\cite{krum,trimmed_mean_median, shejwalkar2021manipulating, li2023experimental, xu2022byzantine, bulyan, fang2020local, cao2021fltrust}. 
These aggregations are specifically designed to mitigate the impact of adversarial clients by identifying and excluding malicious updates based on criteria such as dimension-wise filtering \cite{trimmed_mean_median, bernstein2018signsgd}, vector-wise filtering \cite{krum, bulyan, shejwalkar2021manipulating, li2019rsa}, vector-wise scaling \cite{sun2019can}, knowledge transfer-based techniques \cite{hongyan2019cronus, lin2020ensemble}, singular value decomposition based techniques \cite{diakonikolas2019sever, diakonikolas2017being}, or encoding-based defenses \cite{chen2018draco, konstantinidis2021byzshield, rajput2019detox}. These aggregation rules ensure that only benign updates contribute to the global model. 


\paragraph{Attacks against Byzantine-Robust Aggregations} Despite the robustness of Byzantine-robust aggregation rules, recent studies \cite{fang2020local, shejwalkar2021manipulating, shejwalkar2022back, baruch2019little} have demonstrated that carefully crafted adversarial model updates can bypass them. These attacks exploit vulnerabilities in FL systems by injecting malicious updates that disrupt the aggregation process because they are not filtered out. We refer to these as Byzantine-robust attacks, which can be classified into two categories based on the attacker's knowledge: (1) \textbf{\textit{aggregation-agnostic attacks}} \cite{baruch2019little, shejwalkar2021manipulating} and (2) \textbf{\textit{aggregation-targeted attacks}} \cite{fang2020local, shejwalkar2021manipulating}. In aggregation-agnostic attacks, attackers create malicious updates without knowing the central server's aggregation algorithm, aiming to be effective across various rules by exploiting detection systems' limitations in detecting adversarial patterns. These attacks are challenging because of a lack of knowledge about the server's mechanisms. In contrast, aggregation-targeted attacks use specific knowledge of the server's rules to design updates that bypass defenses. Byzantine-robust attacks differ in strategy and the amount of information accessible to the adversary.FL also allows for various levels of attacker knowledge~\cite{shejwalkar2022back}. In a \textbf{\textit{full knowledge scenario}}~\cite{fang2020local, baruch2019little}, the attacker has access to the code and data of all clients. In a \textbf{\textit{partial knowledge scenario}}~\cite{shejwalkar2021manipulating}, the attacker's access is limited to the adversarial clients only.

\subsection{\textbf{Collusive Model Poisoning Attacks}} These Byzantine-robust attacks \cite{baruch2019little, fang2020local, shejwalkar2021manipulating, shejwalkar2022back} assume that attackers collude to create unified malicious updates to amplify the impact of their attacks. To craft a malicious update, one adversary begins by computing a benign reference aggregate using its local dataset. Then, the adversary generates a malicious update by scaling the benign aggregate in the malicious direction to evade detection by robust aggregations. Once the malicious model update is crafted, the adversary shares it with all other adversarial clients, ensuring that all attackers submit identical (or minimally varied) malicious model updates to the central server. Intuitively, the adversaries are more effective in bypassing the Byzantine robust aggregations if they can coordinate their poisoned updates than if they each acted individually.

\paragraph{Why collusive model poisoning attacks are not practical} Collusion attack has practical challenges. As highlighted in prior work, launching a successful poisoning attack in real-world FL systems typically requires compromising around 20\% of the total clients \cite{shejwalkar2021manipulating, fang2020local}. In a cross-device FL system with 1 million participants, an attacker would need to control about 200,000 devices. They must also ensure reliable communication among these devices to produce consistent malicious updates, often using a botnet or command-and-control (C\&C) server. However, this approach carries the risk of detection by firewalls. Cross-device FL further complicates coordination because clients can join and leave the network unpredictably, making it challenging to maintain synchronized behavior. Additionally, a C\&C server creates a single point of failure: if any part of the system is compromised, the entire attack can fail. Therefore, sustaining reliable, large-scale collusion is impractical in real-world FL deployments, especially in cross-silo FL. Each client represents a trusted organization, such as a hospital or company, with robust security systems and regular monitoring. Breaching multiple secure networks is extremely difficult, and even if successful, coordinating malicious actions would be easily detectable. For these reasons, collusive poisoning attacks are not a realistic threat in cross-silo FL.

\paragraph{Injection of fake clients} Recent research, such as MPAF \cite{cao2022mpaf} and PoisonedFL \cite{xie2024model}, has proposed the use of fake clients in federated learning as a strategy to reduce the need for compromising a large number of genuine clients. These fake clients are entirely controlled by the attacker and allow for the coordination of malicious updates without breaching real user devices. By using virtualization tools such as emulators or virtual machines, an attacker can effectively scale the number of fake clients at a relatively low cost. However, this increases the likelihood of being detected by client verification mechanisms, which can limit its viability in secure or verification-aware federated learning deployments.

\textit{\underline{\textbf{Focus of our work:}}} In this research, we focus on untargeted, non-collusive, aggregation-agnostic Byzantine-robust model poisoning attacks, which pose a significant threat to federated learning systems.

\section{Non-Collusive Poisoning Attacks in Federated Learning}

We introduce a new and realistic adversarial setting in FL called \textbf{\textit{non-collusive poisoning attacks}}. This concept is inspired by the practical challenges of collusive attacks discussed earlier. In this setting, each compromised client independently crafts malicious model updates without coordinating or communicating with other clients or relying on a centralized C\&C server. We define \textbf{\textit{non-collusive poisoning attacks}} as,

\begin{definition}
An attack on an FL system is classified as \textbf{\textit{non-collusive}} if the attacker does not share any information explicitly or implicitly with other compromised clients or any external C\&C server while generating the malicious update.
\end{definition}

To formalize the information boundary for each adversarial client in the \textbf{\textit{non-collusive}} setting, let \( i \in [1..k] \) index a compromised client. Let \( \mathcal{P}_i \) denote its prior knowledge before round \( t \) (e.g., static code, local dataset, or previously observed global models), and let \( \mathcal{K}_i \) denote the total knowledge available to client \( i \) when generating its malicious update \( \phi_i^{m} = \mathcal{A}(\mathcal{K}_i) \), where \( \mathcal{A} \) is the attacker’s local poisoning strategy. We distinguish non-collusive and collusive adversaries based on the presence or absence of shared private information across clients:
\[
\begin{aligned}
\forall i \in& [1 \dots k], \ \forall j \neq i,\notag \\
&(\mathcal{K}_i - \mathcal{P}_i) \cap (\mathcal{K}_j - \mathcal{P}_j) =
\begin{cases}
\emptyset, & \text{if } i \text{ is non-collusive} \\
\neq \emptyset, & \text{if } i \text{ is collusive}
\end{cases}
\end{aligned}
\]

This non-collusive adversary model represents realistic threat scenarios where compromised clients operate independently. This situation is common in malware-based attacks, in which each infected device behaves independently of others. In this context, non-collusive adversaries may initiate either data poisoning or model poisoning attacks, relying solely on the information available to each client. While several previous studies on data poisoning~\cite{mouri23towards, mouri2024data, sun2021data, tolpegin2020data} fit within this non-collusive framework, these attacks often demonstrate limited effectiveness, especially against robust aggregation defenses. In contrast, \textit{a systematic exploration of non-collusive model poisoning has mainly been overlooked in the existing literature}. To the best of our knowledge, this is the first study to investigate non-collusive model poisoning attacks in the context of Byzantine-robust aggregations.

\paragraph{Practicality of Non-Collusive Attacks} Non-collusive attacks are more practical in real-world settings than collusive ones due to their decentralized nature. They do not require coordination among compromised devices, making large-scale attacks more feasible. Devices may be compromised through various means such as malware infections, firmware vulnerabilities, or misconfigured update channels~\cite{valasek2024malware, falowo2024evolving}. Once infected, these devices independently construct malicious updates and submit them to the central server without relying on any peer-to-peer communication. This decentralized approach eliminates the risk of a single point of failure. Even if some compromised devices are detected, others can continue their attacks, thereby ensuring the threat's persistence. This characteristic makes non-collusive attacks more realistic and practical for real-world FL deployments.

\paragraph{Illustrative Threat Scenario} To demonstrate the feasibility of non-collusive  attacks in real-world deployments, we present a hypothetical attack against a production FL system (such as Gboard \cite{hard2018federated}):

\begin{enumerate}
\item An adversary distributes stealthy malware through zero-day exploits, malicious apps, or supply chain vectors to silently infect Android devices. 
\item On each infected device, the malware detects participation in FL by monitoring indicators such as installed packages, FL-related APIs, or known training schedules.
\item Once FL participation is detected, the malware activates a rootkit that intercepts the local FL training pipeline and modifies model updates before transmission.
\item The malware then generates a poisoned model update using only locally available information: its own data, the local model, and the received global model, without any inter-device communication or coordination. 
\item This process operates autonomously on each infected device, enabling large-scale poisoning of the FL process without centralized control or synchronized behavior.
\end{enumerate}

This process mirrors the functionality of real-world malware campaigns such as Triada (modular rooting Trojan), Stuxnet (selective targeting of industrial control systems), and Joker (stealthy exfiltration of data via malicious apps). They illustrate the possibility of stealthy and targeted client-side manipulation in the wild \cite{sharma2025characterization,langner2011stuxnet,shi2020vahunt}. In this scenario, all compromised clients share the same adversarial objective imposed by the malware payload, ensuring aligned attack behavior even without any communication among them.

\section{XFED: A Practical Non-Collusive Model Poisoning Attack}\label{xfed}

We introduced the concept of non-collusive attacks, in which compromised devices operate independently without requiring communication between attackers. To demonstrate this concept, we present XFED, a non-collusive, aggregation-agnostic, untargeted model poisoning attack that functions under a \textit{restrictive} threat model. We will first outline the theoretical threat model of XFED, followed by the mathematical formulation of the XFED attack.

\subsection{Threat Model}

\paragraph{\textbf{Attacker's Goal}} 
Following prior works~\cite{shejwalkar2021manipulating, fang2020local, baruch2019little}, the attacker's objective is to poison the training process of FL such that the final global model produces incorrect predictions on a large portion of test inputs, without targeting specific classes. 


\paragraph{\textbf{Attacker's Capabilities}} 
The attacker can stealthily change the benign model update and send the changed poisonous model update to the central server. It can perform a polynomial time calculation to generate the poisonous model update. However, attackers cannot communicate or coordinate with anything outside of the client device, making the setting fully non-collusive.

\paragraph{\textbf{Attacker's Knowledge}}
We consider a \textit{restrictive} background knowledge scenario, where the attacker's access is limited solely to information on their own device. Specifically, the attacker can access their local training data, local training code, and the sequence of global model updates \(\theta^{(t)}\) received in each round from the central server. The attacker has \textit{no access} to the central server's aggregation rule, global training data, code or updates from other clients.

\subsection{XFED Formulation}
We aim to craft malicious local model updates that push the global model off a benign learning trajectory and degrade its performance. The adversary computes a benign model update \(\phi^{b}\) using local data from the compromised client. To create the malicious update, the attacker selects a perturbation vector \(\psi\) that defines a malicious direction in the gradient space. The final malicious update \( \phi^{m} \) is generated by scaling the benign update in the malicious direction: \( \phi^{m} = \phi^{b} + \mu \psi \), where \( \mu \) is a scaling coefficient that limits the deviation. 
This controlled perturbation allows the attack to bypass Byzantine-robust aggregation defenses while still maximizing global model degradation.

\subsubsection{\textbf{Introducing perturbation vectors}} In the literature, three perturbation vectors ($\psi$) are frequently used for crafting adversarial updates: \textit{inverse sign}, \textit{inverse unit vector}, and \textit{inverse standard deviation} \cite{shejwalkar2021manipulating, shejwalkar2022back, baruch2019little}. We adopt  \textit{inverse sign} (\( \psi_{sgn} \)) and \textit{inverse unit vector} (\( \psi_{uv} \)); but exclude the \textit{inverse standard deviation} direction, as it requires variance across multiple updates, which is infeasible in our proposed non-collusive setting. We denote XFED with the \textit{inverse unit vector} as the perturbation direction as $X_{uv}$, and XFED with the \textit{inverse sign vector} as $X_{sgn}$. We define both the perturbation vectors below.\\
\textbf{Inverse unit vector (\( \psi_{\text{uv}} \))}: $\psi_{uv} = -\frac{\phi^{b}}{\|\phi^{b}\|_2}$\\
\textbf{Inverse sign (\( \psi_{\text{sgn}} \)):} $\psi_{sgn} = -\frac{\text{sign}(\phi^{b})}{\|\text{sign}(\phi^{b})\|_2}$, where \(\text{sign}(\phi^{b})\) returns a vector with elements [+1, -1, 0] based on the sign of each element in \(\phi^{b}\).

\subsubsection{\textbf{Robust Estimation of the Scaling Coefficient}}
\label{mu}

To maximize the impact of \( \phi^{m} \) while avoiding detection as an outlier, the scaling factor \( \mu \) must be chosen carefully. The malicious update \( \phi^{m} \) must steer the global model in the adversarial direction without exceeding the natural variation observed among benign updates \( \phi^{b}_1, \phi^{b}_2, \ldots, \phi^{b}_k \). In other words, it should remain just below the filtering thresholds used by Byzantine-robust aggregators, adapt over time, and appear statistically consistent with honest client behavior. Therefore, our objective is:
\[
\begin{aligned}
\max_{\mu} \mu &= \|\phi^{m} - \phi^{b}\|_2, \text{ s.t. } \mu \text{ not an outlier for the set}\\
&\{\, d \mid d = \|\phi^{b}_i - \phi^{b}_j\|_2,\ \forall\, i,j \in [1,k] \,\}.
\end{aligned}
\]
Prior model poisoning attacks~\cite{shejwalkar2021manipulating, shejwalkar2022back, baruch2019little, fang2020local} compute scaling factors by leveraging benign client updates. However, in the non-collusive setting considered by XFED, adversaries do not have access to benign models. Instead, we propose that the attacker estimate \( \mu \) using the sequence of global models \( (\theta^{(1)}, \theta^{(2)}, \ldots, \theta^{(t)}) \) received up to round \( t \). Our intuition is that, as benign local models progressively converge in each round and the global model is computed as their aggregation, after a sufficient number of rounds, the distances between benign local models become comparable to the distances between successive global models (i.e., the current and previous rounds).

We maintain a list of these distances to compute \( \mu \). To determine a stable and reliable value, we adopt the Median with MAD estimator~\cite{hoaglin2000understanding}. The median is resilient to extreme fluctuations that may arise in early rounds of training or during successful attacks, where the distances between successive global models may not reflect typical benign behavior. Median with MAD provides a robust method for detecting outliers, and we use this threshold to bound \( \mu \). Below, we describe our procedure for computing \( \mu \),


\paragraph{\textbf{Step 1: List of global deltas}} 
For the global model \( \theta^{(t)} \) at round \( t \), we define its delta from the previous round as \( \Delta\theta^{(t)} = \theta^{(t)} - \theta^{(t-1)} \) and maintain a list of all deltas:
\[
\mathcal{H}^{(t)} = \{\, \Delta\theta^{(1)}, \ldots, \Delta\theta^{(t)} \,\}.
\]

\paragraph{\textbf{Step 2: Robust center and scale estimation}}
For each coordinate \(j \in \{1,\ldots,|\Delta\theta^{(t)}|\}\) we compute the coordinate-wise median \(med\) and MAD \(mad\) of \(\mathcal{H}^{(t)}\):
\[
med_j =
  \operatorname*{median}_{\Delta\theta\in\mathcal{H}^{(t)}}(\Delta\theta_j),\qquad
mad_j =
  \operatorname*{median}_{\Delta\theta\in\mathcal{H}^{(t)}}
     \bigl|\Delta\theta_j-med_j\bigr|.
\]

\paragraph{\textbf{Step 3: Determining scaling coefficient based on MAD}}
We then form a robust scale vector \[
s = med + \lambda \cdot mad,
\]
where \(\lambda \ge 0\) controls the aggressiveness of the attack. A recommended maximum value for \(\lambda\) in existing literature~\cite{iglewicz1993volume} is 5.19. However, practically we find any value of \(\lambda \in [2,10]\) has measurable impact. We discuss the choice of \(\lambda\) in more details at Appendix~\ref{lambda-limit}. The final scaling coefficient \(\mu\) is obtained as the \(\ell_2\) norm of this vector:
\[
\mu = \bigl\| s \bigr\|_2
  = \bigl\| med + \lambda \cdot mad \bigr\|_2.
\]
This choice makes \(\mu\) proportional to a robust upper bound on typical coordinate-wise variation of recent global updates.

\subsubsection{\textbf{Final malicious update.}}
Finally, the attacker submits the malicious update \( \phi^{m} = \phi^{b} + \mu \psi \) as the poisoned model. As XFED operates in a non-collusive setting where each adversarial client generates its own malicious update independently, unlike other collusive attacks, XFED does not need random noise injection. However, attackers can introduce lightweight randomness \(\epsilon\) to obscure malicious patterns, such as by using truncated Gaussian jitter for perturbation. The final submitted update then becomes \(\phi^{m} = \phi^{b} + \mu \psi + \epsilon\).

\begin{table*}[tb]
\centering
\caption{Attack impact \( I_\theta \) of the global model under different attacks (20\% malicious clients), aggregation rules, and defenses for different datasets. We assume Fang, LIE, Min-Max, and Min-Sum have access to the model updates on all compromised clients, and Fang further has access to the
aggregation rule, which gives advantages to these attacks. Our proposed \(X_{sgn}\) and \(X_{uv}\) do not have access to any of these knowledge. In each row, we highlight in bold the highest attack impact or the closest to the highest.}
\label{tab:main-results}
\resizebox{\textwidth}{!}{%
\begin{tabular}{@{}l||ccccccccc||ccccccccc@{}}
\toprule
     & \( A_\theta \) & \(X_{sgn}\) & \(X_{uv}\) & Min-Sum & Min-Max & Fang & LIE & MPAF & PoisonedFL & \( A_\theta \) & \(X_{sgn}\) & \(X_{uv}\) & Min-Sum & Min-Max & Fang & LIE & MPAF & PoisonedFL \\ \midrule
 & \multicolumn{9}{c||}{\textbf{(a) Purchase (Cross-silo, 100 FL clients, 500 global iterations \& 3 layer DNN model)}} &      \multicolumn{9}{c}{\textbf{(b) MNIST (Cross-silo, 100 FL clients, 275 global iterations \& 4 layer DNN model)}}      \\ \midrule                                                                       
FedAvg    & 74.82                              & 55.2                            & 67.53                          & 22.01                       & 35.39                       & \textbf{74.35}        & 7.65                    & 65                       & \textbf{72.18}   & 95.02                              & 29.42                           & 81.43                          & 9.39                        & 17.7                        & \textbf{85.22}           & 4.98                    & \textbf{84.92}           & \textbf{83.62}                \\
Median    & 72.14                              & \textbf{37.14}                  & \textbf{36.31}                 & 33.11                       & \textbf{35.46  }            & 7.09                  & 14.85                   & 16.8                     & \textbf{35.7}    & 94.61                              & \textbf{13.73}                           & \textbf{12.21}                 & \textbf{12.5}               & 11.43                       & \textbf{14.31}           & 8                       & 5.15                     & 8.28                          \\
Tr-Mean   & 73.79                              & \textbf{42.71}                  & \textbf{43.43}                 & 33.17                       & 36.1                        & 21.13                 & 14.77                   & 24.24                    & \textbf{43}      & 94.69                              & 14.38                           & 15.19                          & 10.97                       & 13.64                       & \textbf{21.27}           & 6.91                    & 5.55                     & 14                            \\
M-Krum    & 74.54                              & 4.04                            & \textbf{17.48}                 & \textbf{16.85}              & 9.48                        & \textbf{17.55}        & 13.56                   & 3.54                     & 4.67             & 94.57                              & 9.34                            & \textbf{12.5}                  & \textbf{10.32}                       & 9.08                        & \textbf{11.91}           & 7.68                    & 5.58                   & 4.69                          \\
CC        & 74.69                              & 50.48                           & \textbf{68.86}                 & 33.82                       & 48.03                       & -                     & 3.65                    & 4.88                     & 4.71             & 94.08                              & 13.13                           & \textbf{80.1}                  & 13.49                       & 23.57                       & -                      & 4.62                    & 3.32                     & 5.6                          \\
SgnG      & 74.43                              & 50.23                           & \textbf{67.98}                 & 34.8                        & 50.68                       & -                     & 2.94                    & 4.08                     & 3.34             & 95.31                              & 24.57                           & \textbf{81.18}                 & 12.66                       & 25.21                       & -                      & 5.3                     & 3.99                     & 6.87                         \\
FLTrust   & 73.97                              & 2.16                            & \textbf{8.83}                  & \textbf{7.12}               & \textbf{8.55}               & -                     & \textbf{6.99}                    & 4.04                     & 3.18             & 95.49                              & 6.52                           & \textbf{7.35}                  & \textbf{8.2}               & \textbf{8.25}               & -                      & \textbf{7.4}            & 4.98                    & \textbf{7.2}                 \\
FreqFed   & 73.64                              & 3.55                            & \textbf{68.31}                 & 45.38                       & 58.91                       & -                     & 17.24                   & 3.47                     & 4.76             & 94.32                              & 5.95                            & \textbf{80.87}                 & 16.14                       & 35.18                       & -                      & 9.69                    & 5.72                    & 6.79                          \\
FoolsGold & 74.16                              & 15.63                           & \textbf{56.01}                 & 4.41                        & 4.47                        & -                     & 3.69                    & 3.6                      & 2.99             & 94.8                               & \textbf{35.91}                  & \textbf{36.11}                 & 4.48                        & 3.56                        & -                      & 2.49                    & 4.36                     & 3.38                          \\
FLAME     & 74.03                              & \textbf{49.43}                  & \textbf{52.9}                  & 45.75                       & \textbf{52.2 }              & -                     & 26.9                    & 3.23                     & 3.82             & 93.33                              & 3.38                            & \textbf{79.24}                 & 14.93                       & 34.07                        & -                      & 11.34                   & 3.44                     & 3.56                          \\ \midrule

 & \multicolumn{9}{c||}{\textbf{(c) Fashion-MNIST (Cross-silo, 40 FL clients, 250 global iterations \& 3 layer DNN model)}} &      \multicolumn{9}{c}{\textbf{(d) EMNIST (Cross-silo, 200 FL clients, 400 global iterations \& 3 layer DNN model)}}      \\ \midrule

   FedAvg    & 85.19                              & 13.24                           & 43.04                          & 23.92                       & 7.59                        & \textbf{74.35}                   & 6.74                    & 67.44                    & 43.24             & 76.39                              & \textbf{71.24}                  & 25.28                          & 16.22                       & 17.06                       & 26.11                    & 5.11                    & 69.29                    & 65.29

   \\
Median    & 85.18                              & 10.12                           & \textbf{20.22}                          & \textbf{21.44}                       & 7.82                        & \textbf{22.43}           & 14.87                   & 2.5                      & 10.14     & 74.19                              & \textbf{70.96}                  & 21.63                          & 17.07                       & 13.93                       & 4.97                     & 5.95                    & 6.09                     & 19.49                                             

\\
Tr-Mean   & 85.57                              & 10.75                           & 12.22                          & \textbf{15.1}                        & 7.19                        & \textbf{15.2}            & 8.93                    & 3.32                     & 11.76          & 73.79                              & 16.95                           & \textbf{21.66  }                        & 14.55                       & 15.24                       & 10.57                    & 7.06                    & 6.82                     & \textbf{21.89}                              \\
M-Krum    & 85.51                              & 2.73                            & \textbf{10.2}                           & \textbf{10.11}                       & 9.34                        & \textbf{11.2}            & \textbf{10.26}                   & 3.3                      & 3.41             & 74.42                              & 5.79                            & \textbf{21.7}                  & 16.63                       & 19.48                       & 13.26                    & 2.75                    & 7.02                     & 6.86                                 \\
CC        & 85.22                              & 7.14                            & \textbf{42.28}                 & 26.51                       & 11.13                       & -                        & 3.47                    & 3.13                     & 4.09           & 74.66                              & 9.27                            & \textbf{41.17}                 & 19.28                       & 34.44                       & -                        & 3.23                    & 7.12                     & 7.13                                 \\
SgnG      & 85.66                              & 6.86                            & \textbf{42.98}                 & 22.14                       & 9.82                        & -                        & 3.75                    & 3.51                     & 38.13        & 74.42                              & 7.79                            & \textbf{49.26}                 & 18.84                       & 34.52                       & -                        & 1.74                    & 8.03                     & 6                    \\
FLTrust   & 86.32                              & 2.98                            & 3.5                            & 3.56                        & \textbf{5.7}                         & -                        & \textbf{5.825}          & \textbf{5.7}                      & 2.72              & 71.2                               & 2.49                            & 5.72                           & \textbf{11.69}                      & \textbf{12.25}              & -                        & 9.15                    & \textbf{10.12}                    & 2.76                    \\
FreqFed   & 85.68                              & 3.56                            & \textbf{47.13}                 & 25.35                       & 20.42                       & -                        & 8.69                    & 4.66                     & 5.4                   & 70.76                              & 3.43                            & \textbf{65.8}                  & 20.29                       & 36.56                       & -                        & 3.79                    & 3.06                     & 3.48            \\
FoolsGold & 85.69                              & 3.39                            & \textbf{10.22}                 & 3.31                        & 2.81                        & -                        & 3.99                    & 4.18                     & \textbf{10.12}          & 74.39                              & \textbf{6.15}                            & \textbf{6.58}                           & \textbf{6.63}                        & \textbf{6.85}               & -                        & \textbf{6.71}                    & \textbf{6.82}                     & \textbf{5.89}                      \\
FLAME     & 85.64                              & 4.15                            & \textbf{44.17}                 & 23.93                       & 22.04                       & -                        & 9.68                    & 4.2                      & 3.84     & 71.07                              & 4.58                            & \textbf{66.1}                  & 21.03                       & 41.57                       & -                        & 6.56                    & 3.47                     & 4.08                         
 \\ \midrule

 & \multicolumn{9}{c||}{\textbf{(e) Fashion-MNIST (Cross-silo, 40 FL clients, 250 global iterations \& AlexNet model)}} &      \multicolumn{9}{c}{\textbf{(f) CIFAR-10 (Cross-silo, 50 FL clients, 255 global iterations \& AlexNet model)}}      \\ \midrule                                                                       
FedAvg    & 84.98 & 12.16 & 55.27          & 7.51 & 7.59          & \textbf{74.35} & 6.74           & 73.21         & \textbf{73.97} & 86.54 & \textbf{76.54} & \textbf{76.54} & 23.57 & 22.09 & 72.15         & 38.92          & \textbf{76.54} & 62.36          \\
Median    & 84.39 & 8.83  & 18.29          & 6.07 & 7.82          & \textbf{22.43} & 14.87          & 1.53          & \textbf{20.7}  & 82.38 & 20.79          & 26.12          & 19.45 & 20.46 & 34.56         & 18.78          & 8.94           & \textbf{38.16} \\
Tr-Mean   & 84.02 & 7.27  & 18.97          & 5.62 & 7.19          & 15.1           & 8.93           & 1.45          & \textbf{25.42} & 80.4  & 15.23          & \textbf{51.22} & 17.21 & 16.02 & \textbf{52.4} & 38.55          & 8.3            & \textbf{50.85} \\
M-Krum    & 84.25 & 3.08  & \textbf{9.86}  & 6.94 & \textbf{9.34} & \textbf{11.2}           & \textbf{10.26} & 3.37          & 3.96           & 81.56 & 6.69           & \textbf{21.93} & 18.16 & 20.62 & 12.6          & \textbf{22.44} & 6.42           & 3.48           \\
CC        & 83.88 & 2.61  & \textbf{19.94} & 6.67 & 9.79          & -              & 2.13           & 2.85          & 3.33           & 81.04 & 7.15           & \textbf{63.21} & 20.14 & 19.51 & -           & 6.61           & 6.73           & 5.57           \\
SgnG      & 83.6  & 2.19  & \textbf{35.24} & 8.81 & 9.82          & -              & 3.75           & 1.83          & 4.15           & 83.88 & 14.02          & \textbf{28.42} & 16.93 & 16.05 & -           & 4.5            & 11.66          & 7.23           \\
FLTrust   & 83.3  & 0.57  & 1.84           & 1.19 & 1.25          & -              & \textbf{6.27}           & \textbf{8.32} & 2.67           & 80.2  & 4.23           & \textbf{8.77}  & 6.86  & 7.66  & -           & \textbf{8.67}  & \textbf{9.38}  & 6.07           \\
FreqFed   & 83.1  & 15.98 & \textbf{66.39} & 13.1 & 1.75          & -              & 6.11           & 2.43          & 3.24           & 81.34 & 15.11          & \textbf{29.86} & 16.19 & 17.24 & -           & 8.31           & 9.81           & 5.34           \\
FoolsGold & 84.45 & 3.39  & \textbf{4.67}  & 3.36 & 2.41          & -              & \textbf{4.1}            & 2.91          & \textbf{4.06}           & 85.21 & 24.58          & \textbf{53.56} & 12.73 & 11.54 & -           & 10.9           & \textbf{52.18} & 10.55          \\
FLAME     & 83.5  & 5.15  & \textbf{67.54} & 7.17 & 19.9          & -              & 3.94           & 2.56          & 4.03           & 84.25 & 10.28          & \textbf{67.53} & 28.2  & 28.44 & -           & 62.15          & 11.48          & 8.85            \\ \midrule

 & \multicolumn{9}{c||}{\textbf{(g) HAR (Cross-silo, 30 FL clients, 1000 global iterations \& Logistic Regression model)}} &      \multicolumn{9}{c}{\textbf{(h) EMNIST (Cross-device, 50000 FL clients, 400 global iterations \& 3 layer DNN model)}}       \\ \midrule    
 
FedAvg    & 96.98  & 20.12  & 60.12           & 19.51  & 34.13           & 7.43           & 1.08          & \textbf{90.2}  & 57.24           & 71.14 & 50.28          & 60.93          & 14.93          & 41.92          & 50             & 4.85           & \textbf{70.02} & 58.7           \\
Median    & 94.31  & 1.2    & \textbf{12.2}   & 8.39   & 8.24            & 0.89           & 0.42          & \textbf{10.22} & \textbf{10.448} & 70.85 & \textbf{15.18} & \textbf{14.35} & \textbf{15.29} & \textbf{15.01} & \textbf{14.98} & \textbf{15.27} & 4              & \textbf{15.23} \\
Tr-Mean   & 96.79  & 8.71   & \textbf{15.5}   & 10.61  & 11.46           & 3.25           & 2.67          & 10.2           & \textbf{15.49}  & 73.29 & 20.17          & \textbf{21.11} & 17.33          & \textbf{23.48} & \textbf{22.07} & 5.89           & 16.21          & \textbf{22.22} \\
M-Krum    & 96.48  & 8.4    & \textbf{10.46}  & 9.95   & 5.19            & \textbf{12.25} & 2.48          & 0.39           & 0.35            & 75.72 & 10.2           & 15.56          & 18.79          & \textbf{22.39} & \textbf{22.64} & 6.83           & 16.56          & \textbf{20.42} \\
CC        & 97.097 & 13.277 & \textbf{44.257} & 30.497 & \textbf{43.617} & -              & 7.167         & 8.987          & 11.958          & 75.79 & 8.73           & \textbf{65.63} & 28.46          & 52.41          & -              & 2.54           & 8.06           & 8.1            \\
SgnG      & 97.17  & 15.48  & \textbf{44.43}  & 33.63  & \textbf{42.41}  & -              & 1.62          & 2.13           & 40.52           & 75.87 & 9.98           & \textbf{64.52} & 26.99          & 49.66          & -              & 1.43           & 8.82           & 8.4            \\
FLTrust   & 96.05  & 1.94   & 1.82            & 2.09   & 3.25            & -              & \textbf{10.1} & 8.5            & -0.5            & 70.8  & \textbf{5.36}  & 1.81           & \textbf{6.86}  & \textbf{5.83}  & -              & 1.57           & \textbf{5.97}  & 1.98           \\
FreqFed   & 94.81  & 4.72   & \textbf{51.27}  & 30.19  & \textbf{50.54}  & -              & -0.81         & -0.54          & -0.42           & 71.36 & 9.7            & \textbf{66.3}  & 25.7           & 62.45 & -              & 1.91           & 3.68           & 5.08           \\
FoolsGold & 97.21  & 2.17   & \textbf{6.81}   & 2.44   & 4.06            & -              & 2.48          & 3.25           & 2.79            & 74.51 & \textbf{8.15}           & \textbf{10.1}  & \textbf{8.54}           & \textbf{8.95}           & -              & \textbf{9.33}  & \textbf{9.82}  & \textbf{8.58}  \\
FLAME     & 94.93  & 7.59   & \textbf{56.24}  & 41.37  & \textbf{55.04}  & -              & 8.98          & -0.3           & 0.08            & 72.97 & 7.08           & \textbf{68}    & 33.86          & \textbf{68}    & -              & 3.34           & 6.1            &  6.09            
 \\ \midrule

\end{tabular}%
}
\end{table*}

\section{Evaluation}\label{experimental}

\subsection{Experimental Setup}

\paragraph{FL Settings \& Datasets} We use datasets from different domains, like four image classification benchmarks (MNIST~\cite{deng2012mnist}, EMNIST~\cite{cohen2017emnist}, Fashion-MNIST~\cite{xiao2017fashion}, CIFAR-10~\cite{cifar10}), one customer transaction behavior dataset (Purchase-100~\cite{wen2018customer}), and one human activity recognition dataset (HAR)~\cite{anguita2013public}. Unless specified, all datasets are independent and identically distributed (iid) except HAR. Detailed descriptions of the datasets and model architectures are in the Appendix ~\ref{dataset}. We consider cross-silo FL settings unless noted otherwise. For our attack XFED, we set \textbf{\(\lambda = 4\)}.

\paragraph{Measurement Metrics} Let \( A_\theta \) and \( A^*_\theta \) denote the average accuracy of the final 10\% of training rounds in both the no-attack and the under-attack settings. We define the attack impact \( I_\theta \) as \( I_\theta = A_\theta - A^*_\theta \).

\paragraph{Baseline Aggregations, Defenses, \& Attacks}
We evaluate the performance of our proposed attacks (\(X_{sgn}\) \& \(X_{uv}\)) against most popular aggregation  FedAvg~\cite{fedAvg}, five byzantine-robust aggregations (Multi-Krum~\cite{krum}, Trimmed-Mean(Tr-Mean)~\cite{trimmed_mean_median}, Median~\cite{trimmed_mean_median}, Clipped-Clustering (CC)~\cite{li2023experimental}, and SignGuard(SgnG)~\cite{xu2022byzantine}), and four state-of-the-art defenses (FLTrust \cite{cao2021fltrust}, FLAME \cite{nguyen2022flame}, FoolsGold \cite{fung2018mitigating}, and FreqFed \cite{fereidoonifreqfed}). We compare our attacks with other attacks that aim to meet the practicality requirements defined by Shejwalkar et al.~\cite{shejwalkar2022back}, MPAF~\cite{cao2022mpaf}, and PoisonedFL~\cite{xie2024model}, which use the injection of fake clients into the FL system. For a complete benchmark, we also compare with other non-practical state-of-the-art attacks, including three non-aggregation-agnostic attacks (LIE~\cite{baruch2019little}, Min-Max~\cite{shejwalkar2021manipulating}, and Min-Sum~\cite{shejwalkar2021manipulating}) and an aggregation-targeted attack, Fang~\cite{fang2020local}. Due to space constraints, detailed descriptions of these attacks, aggregations, and defenses are provided in Appendix~\ref {baseline-attacks-appendix}, \ref{aggregation-appendix}, and \ref{defenses-appendix}. Unless specified, we assume that 20\% of clients are malicious in all our experiments. For MPAF and PoisonedFL, when we refer to \% of malicious clients, we mean the \% of fake clients injected.

\subsection{Main Results} \label{results-analysis}

In this section, we compare the performance of \(X_{sgn}\) and \(X_{uv}\) with state-of-the-art model poisoning attacks. The results are presented in Table~\ref{tab:main-results}, where the \( A_\theta \) column indicates ``No Attack" setting, while the remaining columns report the attack impact \( I_\theta \) for different attacks. For MPAF and PoisonedFL ``No Attack" means no fake clients are injected and for the other attacks it means no clients are compromised.

\input{xfed_results_modified}

\input{noniid}

\paragraph{XFED outperforms existing attacks} As shown in Table~\ref{tab:main-results}, \textbf{our proposed \(X_{uv}\) attack outperforms all state-of-the-art model poisoning attacks for all the combinations of aggregation rules, dataset, and model architecture by substantial margins in almost all cases despite the absence of client-to-client communication}. For instance, on the MNIST dataset, \(X_{uv}\) attack is 2.29× to 14.135× more impactful than the state-of-the-art attacks. LIE is ineffective in most cases when a defense is deployed; while MPAF and PoisonedFL are effective to some extent against some defenses, e.g., Median, Tr-Mean, and FoolsGold, they are ineffective against others, e.g., CC, SgnG, and FLAME. Our proposed \(X_{uv}\) substantially outperforms Min-Sum, Min-Max, LIE, and MPAF in almost all cases. For instance, on the Fashion-MNIST dataset, \(X_{uv}\) achieves 3.39x higher impact than Min-Max at attacking FLAME. We detail why \(X_{uv}\) outperforms existing attacks in Appendix~\ref{effectiveness_xfed}. One important observation is that the Fang attack sometimes outperforms our attacks when the aggregation is Multi-Krum, and in some cases Tr-Mean by a small margin. This is because Fang's attack is aggregation-targeted and specifically tailored for Krum and Tr-Mean aggregation, so the performance gap is expected.

\paragraph{XFED breaks state-of-the-art FL defenses} We observe that \textbf{none of the state-of-the-art aggregations or defenses (except for FLtrust) can effectively mitigate our proposed \(X_{uv}\) attack for any combination of aggregation rules, datasets, or model architectures}. Specifically, \(X_{uv}\) increases \( I_\theta \) by 7.35 to 81.18 across the MNIST dataset. Our results show that existing FL defenses are not robust, even if an attacker does not know the defense deployed on the server or any information about the other clients (benign or compromised). Our attacks and other state-of-the-art attacks work exceptionally well against FedAvg, as it is not Byzantine-robust and fails to filter out malicious updates. We detail why \(X_{uv}\) breaks FL defenses in Appendix \ref{outperform_xfed}. However, FLTrust mitigates the \(X_{uv}\) attack and other attacks by using a small trusted root dataset in the server to compute a reference gradient that allows it to filter updates that deviate from the expected direction. While Table~\ref{tab:main-results} suggests FoolsGold is an effective defense, its performance varies significantly across datasets. Fig.~\ref{impact-mnist} reveals its unreliability, as it performs poorly with few attackers but improves with more malicious clients. However, its effectiveness drops again at higher attack ratios, underscoring its instability.



\paragraph{Impact of cross-device setting} We also compare the performance of \(X_{sgn}\) and \(X_{uv}\) in cross-device settings for EMNIST (Table~\ref{tab:main-results}) and Purchase (Appendix~\ref{cross-device-appendix}) datasets. In this setting, we simulate an FL system of 50,000 clients, and in each training round, the server selects a random subset of 1\% (500 clients) to participate in training. Similar to cross-silo setting, \textbf{\(X_{uv}\) outperforms all other attacks for all the combinations of aggregations and defenses (except FLTrust) by a substantial margin in almost all cases.}

\subsection{Impact of FL Settings}

\paragraph{Impact of the percentage of malicious clients} Fig.~\ref{impact-mnist} shows the impact of model poisoning attacks on the learnt global model as the percentage of malicious clients in FL varies from 1\% to 30\% for the MNIST datasets. \textbf{\(X_{uv}\) outperforms all other attacks in most settings and across all percentages of malicious clients}. First, we observe that our proposed \(X_{uv}\) attack is more effective; the attack impact increases as the fraction of malicious clients is increased. Second, even with only 5\% of malicious clients \(X_{uv}\) can already break CC, SgnG FoolsGold, FLAME, and FreqFred. However, existing attacks cannot reach such effectiveness until 20\% of clients are malicious. We note that as the percentage of malicious clients increases, the impact of our attacks and the differences between them and existing ones become more significant across all aggregation rules, in most cases. Due to limited space, we show the impact of the percentage of malicious clients on MNIST(Fed-Avg \& Median) and the Purchase dataset in Appendix~\ref{purchase-all-attacker}.

\paragraph{Impact of the degree of non-IID} In our experiments, we use a parameter p to control the degree of non-IID among the clients' local training data, where $0.1 \leq p \leq 1$ \cite{fang2020local}. A larger value of p indicates a higher degree of non-IID data. Fig.~\ref{fig:non-iid} shows the impact of different attacks on the global model as p increases from 0.1 to 0.9 on the Purchase dataset. We detail the description of simulating non-IID data distribution and the impact of the degree of non-IID on the Purchase (Fed-Avg \& Median) and MNIST datasets in Appendix~\ref{noniid-appendix}. We observe that \(X_{uv}\) consistently achieves high attack impact across different settings of p under different defenses. However, we also note that as the degree of non-IID increases, the global model accuracy, \( A_\theta \), decreases (as shown in the graph), and consequently, the attack impact decreases.

\paragraph{\textbf{Impact of perturbation vectors}} We find that, for a given dataset, model, and aggregation, varying the perturbation (\(\psi_{\text{sgn}}\) and \(\psi_{\text{uv}}\)) significantly changes the impact of our attacks. Overall, the \(\psi_{\text{uv}}\) tends to produce stronger attack impacts and in Appendix~\ref{disc-pert}, we discuss possible reasons.

Due to space limitations, we provide additional results including the effects of varying the percentage of malicious clients in the cross-device setting, the impact of different neural network architectures, and the influence of various attack parameters in Appendix \ref{purchase-all-attacker}, \ref{diff-neural-network-appendix}, and \ref{attack-parameter-appendix}.

\section{Practical Defense Recommendations}\label{defense}

We disagree with the conclusion of Shejwalkar et al.~\cite{shejwalkar2022back} that FedAvg alone is sufficient to protect real-world FL systems from poisoning attacks. Our experiments show that \textbf{FLTrust} is the most consistent and effective defense across all evaluated attacks, even with 30\% adversarial clients. When a small, trusted root dataset can be collected, we recommend FLTrust for practical FL deployments. However, its performance is highly sensitive to the class distribution of the root dataset. In our experiments (Appendix~\ref{fltrust-root-bias}), on the Purchase dataset, increasing the bias from 0.1 to 0.7 leads to a substantial accuracy drop (from 73.91\% to 61.9\%). When a root dataset is unavailable, we recommend the Multi-Krum, which is effective for lower adversarial participation rates (i.e., fewer than 20\% malicious clients).

\section{Conclusions}\label{conclusion}

In this work, we introduce the concept of non-collusive attacks and propose a more practical threat model for poisoning in FL. This model assumes that malicious clients act independently, without coordination or shared information. We show that realistic adversaries can compromise a modest fraction of clients via common attack vectors, such as malware, and launch highly effective, stealthy poisoning attacks. To demonstrate this feasibility, we present the XFED framework as a representative non-collusive attack model. Our results reveal that an attacker can substantially degrade the global model even without knowledge of other adversarial or benign clients, or of the aggregation rules and defenses deployed at the server. Overall, XFED exposes a critical vulnerability in current FL systems and highlights the urgent need to develop more robust defenses.

{
    \small
    \bibliographystyle{ieeenat_fullname}
    \bibliography{main}

\begin{thebibliography}{59}
\providecommand{\natexlab}[1]{#1}
\providecommand{\url}[1]{\texttt{#1}}
\expandafter\ifx\csname urlstyle\endcsname\relax
  \providecommand{\doi}[1]{doi: #1}\else
  \providecommand{\doi}{doi: \begingroup \urlstyle{rm}\Url}\fi

\bibitem[Anguita et~al.(2013)Anguita, Ghio, Oneto, Parra, Reyes-Ortiz, et~al.]{anguita2013public}
Davide Anguita, Alessandro Ghio, Luca Oneto, Xavier Parra, Jorge~Luis Reyes-Ortiz, et~al.
\newblock A public domain dataset for human activity recognition using smartphones.
\newblock In \emph{Esann}, pages 3--4, 2013.

\bibitem[Bagdasaryan et~al.(2020)Bagdasaryan, Veit, Hua, Estrin, and Shmatikov]{bagdasaryan2020backdoor}
Eugene Bagdasaryan, Andreas Veit, Yiqing Hua, Deborah Estrin, and Vitaly Shmatikov.
\newblock How to backdoor federated learning.
\newblock In \emph{International Conference on Artificial Intelligence and Statistics}, pages 2938--2948. PMLR, 2020.

\bibitem[Baruch et~al.(2019)Baruch, Baruch, and Goldberg]{baruch2019little}
Gilad Baruch, Moran Baruch, and Yoav Goldberg.
\newblock A little is enough: Circumventing defenses for distributed learning.
\newblock \emph{Advances in Neural Information Processing Systems}, 32, 2019.

\bibitem[Bernstein et~al.(2018)Bernstein, Zhao, Azizzadenesheli, and Anandkumar]{bernstein2018signsgd}
Jeremy Bernstein, Jiawei Zhao, Kamyar Azizzadenesheli, and Anima Anandkumar.
\newblock signsgd with majority vote is communication efficient and fault tolerant.
\newblock \emph{arXiv preprint arXiv:1810.05291}, 2018.

\bibitem[Bhagoji et~al.(2019)Bhagoji, Chakraborty, Mittal, and Calo]{bhagoji2019analyzing}
Arjun~Nitin Bhagoji, Supriyo Chakraborty, Prateek Mittal, and Seraphin Calo.
\newblock Analyzing federated learning through an adversarial lens.
\newblock In \emph{International Conference on Machine Learning}, pages 634--643. PMLR, 2019.

\bibitem[Blanchard et~al.(2017{\natexlab{a}})Blanchard, El~Mhamdi, Guerraoui, and Stainer]{blanchard2017machine}
Peva Blanchard, El~Mahdi El~Mhamdi, Rachid Guerraoui, and Julien Stainer.
\newblock Machine learning with adversaries: Byzantine tolerant gradient descent.
\newblock \emph{Advances in neural information processing systems}, 30, 2017{\natexlab{a}}.

\bibitem[Blanchard et~al.(2017{\natexlab{b}})Blanchard, El~Mhamdi, Guerraoui, and Stainer]{krum}
Peva Blanchard, El~Mahdi El~Mhamdi, Rachid Guerraoui, and Julien Stainer.
\newblock Machine learning with adversaries: Byzantine tolerant gradient descent.
\newblock \emph{Advances in neural information processing systems}, 30, 2017{\natexlab{b}}.

\bibitem[Cao and Gong(2022)]{cao2022mpaf}
Xiaoyu Cao and Neil~Zhenqiang Gong.
\newblock Mpaf: Model poisoning attacks to federated learning based on fake clients.
\newblock In \emph{Proceedings of the IEEE/CVF Conference on Computer Vision and Pattern Recognition}, pages 3396--3404, 2022.

\bibitem[Cao et~al.(2021)Cao, Fang, Liu, and Gong]{cao2021fltrust}
Xiaoyu Cao, Minghong Fang, Jia Liu, and Neil~Zhenqiang Gong.
\newblock Fltrust: Byzantine-robust federated learning via trust bootstrapping.
\newblock In \emph{ISOC Network and Distributed System Security Symposium (NDSS)}, 2021.

\bibitem[Chen et~al.(2018)Chen, Wang, Charles, and Papailiopoulos]{chen2018draco}
Lingjiao Chen, Hongyi Wang, Zachary Charles, and Dimitris Papailiopoulos.
\newblock Draco: Byzantine-resilient distributed training via redundant gradients.
\newblock In \emph{International Conference on Machine Learning}, pages 903--912. PMLR, 2018.

\bibitem[Cohen et~al.(2017)Cohen, Afshar, Tapson, and Van~Schaik]{cohen2017emnist}
Gregory Cohen, Saeed Afshar, Jonathan Tapson, and Andre Van~Schaik.
\newblock Emnist: Extending mnist to handwritten letters.
\newblock In \emph{2017 international joint conference on neural networks (IJCNN)}, pages 2921--2926. IEEE, 2017.

\bibitem[Deng(2012)]{deng2012mnist}
Li Deng.
\newblock The mnist database of handwritten digit images for machine learning research [best of the web].
\newblock \emph{IEEE signal processing magazine}, 29\penalty0 (6):\penalty0 141--142, 2012.

\bibitem[Diakonikolas et~al.(2017)Diakonikolas, Kamath, Kane, Li, Moitra, and Stewart]{diakonikolas2017being}
Ilias Diakonikolas, Gautam Kamath, Daniel~M Kane, Jerry Li, Ankur Moitra, and Alistair Stewart.
\newblock Being robust (in high dimensions) can be practical.
\newblock In \emph{International Conference on Machine Learning}, pages 999--1008. PMLR, 2017.

\bibitem[Diakonikolas et~al.(2019)Diakonikolas, Kamath, Kane, Li, Steinhardt, and Stewart]{diakonikolas2019sever}
Ilias Diakonikolas, Gautam Kamath, Daniel Kane, Jerry Li, Jacob Steinhardt, and Alistair Stewart.
\newblock Sever: A robust meta-algorithm for stochastic optimization.
\newblock In \emph{International Conference on Machine Learning}, pages 1596--1606. PMLR, 2019.

\bibitem[Estenssoro(2024)]{valasek2024malware}
Jessica~Valasek Estenssoro.
\newblock Malware and virus statistics 2024: The trends you need to know about.
\newblock \url{https://www.avg.com/en/signal/malware-statistics}, 2024.
\newblock Published on AVG Antivirus' website.

\bibitem[Falowo et~al.(2024)Falowo, Ozer, Li, and Abdo]{falowo2024evolving}
Olufunsho~I. Falowo, Murat Ozer, Chengcheng Li, and Jacques~Bou Abdo.
\newblock Evolving malware and ddos attacks: Decadal longitudinal study.
\newblock \emph{IEEE Access}, 12:\penalty0 39221--39237, 2024.

\bibitem[Fang et~al.(2020)Fang, Cao, Jia, and Gong]{fang2020local}
Minghong Fang, Xiaoyu Cao, Jinyuan Jia, and Neil Gong.
\newblock Local model poisoning attacks to $\{$Byzantine-Robust$\}$ federated learning.
\newblock In \emph{29th USENIX Security Symposium (USENIX Security 20)}, pages 1605--1622, 2020.

\bibitem[Fereidooni et~al.(2025)Fereidooni, Pegoraro, Rieger, Dmitrienko, and Sadeghi]{fereidoonifreqfed}
Hossein Fereidooni, Alessandro Pegoraro, Phillip Rieger, Alexandra Dmitrienko, and Ahmad-Reza Sadeghi.
\newblock Freqfed: A frequency analysis-based approach for mitigating poisoning attacks in federated learning.
\newblock 2025.

\bibitem[Fung et~al.(2018)Fung, Yoon, and Beschastnikh]{fung2018mitigating}
Clement Fung, Chris~JM Yoon, and Ivan Beschastnikh.
\newblock Mitigating sybils in federated learning poisoning.
\newblock \emph{arXiv preprint arXiv:1808.04866}, 2018.

\bibitem[Guerraoui et~al.(2018)Guerraoui, Rouault, et~al.]{bulyan}
Rachid Guerraoui, S{\'e}bastien Rouault, et~al.
\newblock The hidden vulnerability of distributed learning in byzantium.
\newblock In \emph{International Conference on Machine Learning}, pages 3521--3530. PMLR, 2018.

\bibitem[Hard et~al.(2018)Hard, Rao, Mathews, Ramaswamy, Beaufays, Augenstein, Eichner, Kiddon, and Ramage]{hard2018federated}
Andrew Hard, Kanishka Rao, Rajiv Mathews, Swaroop Ramaswamy, Fran{\c{c}}oise Beaufays, Sean Augenstein, Hubert Eichner, Chlo{\'e} Kiddon, and Daniel Ramage.
\newblock Federated learning for mobile keyboard prediction.
\newblock \emph{arXiv preprint arXiv:1811.03604}, 2018.

\bibitem[Hoaglin et~al.(2000)Hoaglin, Mosteller, and Tukey]{hoaglin2000understanding}
David~C Hoaglin, Frederick Mosteller, and John~W Tukey.
\newblock \emph{Understanding robust and exploratory data analysis}.
\newblock John Wiley \& Sons, 2000.

\bibitem[Hongyan et~al.(2019)Hongyan, Virat, Reza, and Amir]{hongyan2019cronus}
Chang Hongyan, Shejwalkar Virat, Shokri Reza, and Houmansadr Amir.
\newblock Cronus: Robust and heterogeneous collaborative learning with black-box knowledge transfer.
\newblock \emph{arXiv preprint arXiv:1912.11279}, 2019.

\bibitem[Huang et~al.(2022)Huang, Huang, and Liu]{huang2022crosssilo}
Chao Huang, Jianwei Huang, and Xin Liu.
\newblock Cross-silo federated learning: Challenges and opportunities.
\newblock \emph{arXiv preprint arXiv:2206.12949}, 2022.

\bibitem[Huang et~al.(2021)Huang, Chu, Zhou, Wang, Liu, Pei, and Zhang]{huangCross-silo}
Yutao Huang, Lingyang Chu, Zirui Zhou, Lanjun Wang, Jiangchuan Liu, Jian Pei, and Yong Zhang.
\newblock Personalized cross-silo federated learning on non-iid data.
\newblock \emph{Proceedings of the AAAI Conference on Artificial Intelligence}, 35\penalty0 (9):\penalty0 7865--7873, 2021.

\bibitem[Iglewicz and Hoaglin(1993)]{iglewicz1993volume}
Boris Iglewicz and David~C Hoaglin.
\newblock \emph{Volume 16: how to detect and handle outliers}.
\newblock Quality Press, 1993.

\bibitem[Jagielski et~al.(2018)Jagielski, Oprea, Biggio, Liu, Nita-Rotaru, and Li]{jagielski2018manipulating}
Matthew Jagielski, Alina Oprea, Battista Biggio, Chang Liu, Cristina Nita-Rotaru, and Bo Li.
\newblock Manipulating machine learning: Poisoning attacks and countermeasures for regression learning.
\newblock In \emph{2018 IEEE Symposium on Security and Privacy (SP)}, pages 19--35. IEEE, 2018.

\bibitem[Karimireddy et~al.(2021)Karimireddy, Jaggi, Kale, Mohri, Reddi, Stich, and Suresh]{cross-device}
Sai~Praneeth Karimireddy, Martin Jaggi, Satyen Kale, Mehryar Mohri, Sashank Reddi, Sebastian~U Stich, and Ananda~Theertha Suresh.
\newblock Breaking the centralized barrier for cross-device federated learning.
\newblock 34:\penalty0 28663--28676, 2021.

\bibitem[Kone{\v{c}}n{\`y} et~al.(2016{\natexlab{a}})Kone{\v{c}}n{\`y}, McMahan, Ramage, and Richt{\'a}rik]{konevcny2016federated}
Jakub Kone{\v{c}}n{\`y}, H~Brendan McMahan, Daniel Ramage, and Peter Richt{\'a}rik.
\newblock Federated optimization: Distributed machine learning for on-device intelligence.
\newblock \emph{arXiv preprint arXiv:1610.02527}, 2016{\natexlab{a}}.

\bibitem[Kone{\v{c}}n{\`y} et~al.(2016{\natexlab{b}})Kone{\v{c}}n{\`y}, McMahan, Yu, Richt{\'a}rik, Suresh, and Bacon]{konevcny2016federatedStrategies}
Jakub Kone{\v{c}}n{\`y}, H~Brendan McMahan, Felix~X Yu, Peter Richt{\'a}rik, Ananda~Theertha Suresh, and Dave Bacon.
\newblock Federated learning: Strategies for improving communication efficiency.
\newblock \emph{arXiv preprint arXiv:1610.05492}, 2016{\natexlab{b}}.

\bibitem[Konstantinidis and Ramamoorthy(2021)]{konstantinidis2021byzshield}
Konstantinos Konstantinidis and Aditya Ramamoorthy.
\newblock Byzshield: An efficient and robust system for distributed training.
\newblock \emph{Proceedings of Machine Learning and Systems}, 3:\penalty0 812--828, 2021.

\bibitem[Krizhevsky et~al.(2009)Krizhevsky, Hinton, et~al.]{cifar10}
Alex Krizhevsky, Geoffrey Hinton, et~al.
\newblock Learning multiple layers of features from tiny images, 2009.

\bibitem[Langner(2011)]{langner2011stuxnet}
Ralph Langner.
\newblock Stuxnet: Dissecting a cyberwarfare weapon.
\newblock \emph{IEEE Security \& Privacy}, 9\penalty0 (3):\penalty0 49--51, 2011.

\bibitem[Li et~al.(2019)Li, Xu, Chen, Giannakis, and Ling]{li2019rsa}
Liping Li, Wei Xu, Tianyi Chen, Georgios~B Giannakis, and Qing Ling.
\newblock Rsa: Byzantine-robust stochastic aggregation methods for distributed learning from heterogeneous datasets.
\newblock In \emph{Proceedings of the AAAI conference on artificial intelligence}, pages 1544--1551, 2019.

\bibitem[Li et~al.(2023)Li, Ngai, and Voigt]{li2023experimental}
Shenghui Li, Edith C-H Ngai, and Thiemo Voigt.
\newblock An experimental study of byzantine-robust aggregation schemes in federated learning.
\newblock \emph{IEEE Transactions on Big Data}, 2023.

\bibitem[Lin et~al.(2020)Lin, Kong, Stich, and Jaggi]{lin2020ensemble}
Tao Lin, Lingjing Kong, Sebastian~U Stich, and Martin Jaggi.
\newblock Ensemble distillation for robust model fusion in federated learning.
\newblock \emph{Advances in neural information processing systems}, 33:\penalty0 2351--2363, 2020.

\bibitem[Mahloujifar et~al.(2019)Mahloujifar, Mahmoody, and Mohammed]{mahloujifar2019universal}
Saeed Mahloujifar, Mohammad Mahmoody, and Ameer Mohammed.
\newblock Universal multi-party poisoning attacks.
\newblock In \emph{International Conference on Machine Learning}, pages 4274--4283. PMLR, 2019.

\bibitem[McMahan et~al.(2017{\natexlab{a}})McMahan, Moore, Ramage, Hampson, and Arcas]{fedAvg}
Brendan McMahan, Eider Moore, Daniel Ramage, Seth Hampson, and Blaise Aguera~y Arcas.
\newblock {Communication-Efficient Learning of Deep Networks from Decentralized Data}.
\newblock In \emph{Proceedings of the 20th International Conference on Artificial Intelligence and Statistics}, pages 1273--1282. PMLR, 2017{\natexlab{a}}.

\bibitem[McMahan et~al.(2017{\natexlab{b}})McMahan, Moore, Ramage, Hampson, and y~Arcas]{mcmahan2017communication}
Brendan McMahan, Eider Moore, Daniel Ramage, Seth Hampson, and Blaise~Aguera y Arcas.
\newblock Communication-efficient learning of deep networks from decentralized data.
\newblock In \emph{Artificial intelligence and statistics}, pages 1273--1282. PMLR, 2017{\natexlab{b}}.

\bibitem[Mouri et~al.(2023)Mouri, Ridowan, and Adnan]{mouri23towards}
Israt~Jahan Mouri, Muhammad Ridowan, and Muhammad~Abdullah Adnan.
\newblock Towards poisoning of federated support vector machines with data poisoning attacks.
\newblock In \emph{Proceedings of the 13th International Conference on Cloud Computing and Services Science - CLOSER,}, pages 24--33. INSTICC, SciTePress, 2023.

\bibitem[Mouri et~al.(2024)Mouri, Ridowan, and Adnan]{mouri2024data}
Israt~Jahan Mouri, Muhammad Ridowan, and Muhammad~Abdullah Adnan.
\newblock Data poisoning attacks and mitigation strategies on federated support vector machines.
\newblock \emph{SN Computer Science}, 5\penalty0 (2):\penalty0 241, 2024.

\bibitem[Mu{\~n}oz-Gonz{\'a}lez et~al.(2017)Mu{\~n}oz-Gonz{\'a}lez, Biggio, Demontis, Paudice, Wongrassamee, Lupu, and Roli]{munoz2017towards}
Luis Mu{\~n}oz-Gonz{\'a}lez, Battista Biggio, Ambra Demontis, Andrea Paudice, Vasin Wongrassamee, Emil~C Lupu, and Fabio Roli.
\newblock Towards poisoning of deep learning algorithms with back-gradient optimization.
\newblock In \emph{Proceedings of the 10th ACM workshop on artificial intelligence and security}, pages 27--38, 2017.

\bibitem[Nguyen et~al.(2022)Nguyen, Rieger, De~Viti, Chen, Brandenburg, Yalame, M{\"o}llering, Fereidooni, Marchal, Miettinen, et~al.]{nguyen2022flame}
Thien~Duc Nguyen, Phillip Rieger, Roberta De~Viti, Huili Chen, Bj{\"o}rn~B Brandenburg, Hossein Yalame, Helen M{\"o}llering, Hossein Fereidooni, Samuel Marchal, Markus Miettinen, et~al.
\newblock $\{$FLAME$\}$: Taming backdoors in federated learning.
\newblock In \emph{31st USENIX Security Symposium (USENIX Security 22)}, pages 1415--1432, 2022.

\bibitem[Rajput et~al.(2019)Rajput, Wang, Charles, and Papailiopoulos]{rajput2019detox}
Shashank Rajput, Hongyi Wang, Zachary Charles, and Dimitris Papailiopoulos.
\newblock Detox: A redundancy-based framework for faster and more robust gradient aggregation.
\newblock \emph{Advances in Neural Information Processing Systems}, 32, 2019.

\bibitem[Rehman et~al.(2021)Rehman, Dirir, Salah, Damiani, and Svetinovic]{rehman-cross-device}
Muhammad Habib~ur Rehman, Ahmed~Mukhtar Dirir, Khaled Salah, Ernesto Damiani, and Davor Svetinovic.
\newblock Trustfed: A framework for fair and trustworthy cross-device federated learning in iiot.
\newblock \emph{IEEE Transactions on Industrial Informatics}, 17\penalty0 (12):\penalty0 8485--8494, 2021.

\bibitem[Sharma and Rattan(2025)]{sharma2025characterization}
Tejpal Sharma and Dhavleesh Rattan.
\newblock Characterization of android malwares and their families.
\newblock \emph{ACM Comput. Surv.}, 57\penalty0 (5), 2025.

\bibitem[Shejwalkar and Houmansadr(2021)]{shejwalkar2021manipulating}
Virat Shejwalkar and Amir Houmansadr.
\newblock Manipulating the byzantine: Optimizing model poisoning attacks and defenses for federated learning.
\newblock In \emph{NDSS}, 2021.

\bibitem[Shejwalkar et~al.(2022)Shejwalkar, Houmansadr, Kairouz, and Ramage]{shejwalkar2022back}
Virat Shejwalkar, Amir Houmansadr, Peter Kairouz, and Daniel Ramage.
\newblock Back to the drawing board: A critical evaluation of poisoning attacks on production federated learning.
\newblock In \emph{IEEE Symposium on Security and Privacy}, 2022.

\bibitem[Shi et~al.(2020)Shi, Ming, Fu, Peng, Xu, Gao, and Pan]{shi2020vahunt}
Luman Shi, Jiang Ming, Jianming Fu, Guojun Peng, Dongpeng Xu, Kun Gao, and Xuanchen Pan.
\newblock Vahunt: Warding off new repackaged android malware in app-virtualization's clothing.
\newblock In \emph{Proceedings of the 2020 ACM SIGSAC conference on computer and communications security}, page 535–549, New York, NY, USA, 2020. Association for Computing Machinery.

\bibitem[Sun et~al.(2021)Sun, Cong, Dong, Wang, Lyu, and Liu]{sun2021data}
Gan Sun, Yang Cong, Jiahua Dong, Qiang Wang, Lingjuan Lyu, and Ji Liu.
\newblock Data poisoning attacks on federated machine learning.
\newblock \emph{IEEE Internet of Things Journal}, 2021.

\bibitem[Sun et~al.(2019)Sun, Kairouz, Suresh, and McMahan]{sun2019can}
Ziteng Sun, Peter Kairouz, Ananda~Theertha Suresh, and H~Brendan McMahan.
\newblock Can you really backdoor federated learning?
\newblock \emph{arXiv preprint arXiv:1911.07963}, 2019.

\bibitem[Tolpegin et~al.(2020)Tolpegin, Truex, Gursoy, and Liu]{tolpegin2020data}
Vale Tolpegin, Stacey Truex, Mehmet~Emre Gursoy, and Ling Liu.
\newblock Data poisoning attacks against federated learning systems.
\newblock In \emph{European Symposium on Research in Computer Security}, pages 480--501. Springer, 2020.

\bibitem[Wang et~al.(2020)Wang, Sreenivasan, Rajput, Vishwakarma, Agarwal, Sohn, Lee, and Papailiopoulos]{wang2020attack}
Hongyi Wang, Kartik Sreenivasan, Shashank Rajput, Harit Vishwakarma, Saurabh Agarwal, Jy-yong Sohn, Kangwook Lee, and Dimitris Papailiopoulos.
\newblock Attack of the tails: Yes, you really can backdoor federated learning.
\newblock \emph{Advances in Neural Information Processing Systems}, 33:\penalty0 16070--16084, 2020.

\bibitem[Wen et~al.(2018)Wen, Yeh, Tsai, Peng, and Shuai]{wen2018customer}
Yu-Ting Wen, Pei-Wen Yeh, Tzu-Hao Tsai, Wen-Chih Peng, and Hong-Han Shuai.
\newblock Customer purchase behavior prediction from payment datasets.
\newblock In \emph{Proceedings of the Eleventh ACM International Conference on Web Search and Data Mining}, pages 628--636, 2018.

\bibitem[Xiao et~al.(2017)Xiao, Rasul, and Vollgraf]{xiao2017fashion}
Han Xiao, Kashif Rasul, and Roland Vollgraf.
\newblock Fashion-mnist: a novel image dataset for benchmarking machine learning algorithms.
\newblock \emph{arXiv preprint arXiv:1708.07747}, 2017.

\bibitem[Xie et~al.(2020)Xie, Koyejo, and Gupta]{xie2020fall}
Cong Xie, Oluwasanmi Koyejo, and Indranil Gupta.
\newblock Fall of empires: Breaking byzantine-tolerant sgd by inner product manipulation.
\newblock In \emph{Uncertainty in Artificial Intelligence}, pages 261--270. PMLR, 2020.

\bibitem[Xie et~al.(2025)Xie, Fang, and Gong]{xie2024model}
Yueqi Xie, Minghong Fang, and Neil~Zhenqiang Gong.
\newblock Model poisoning attacks to federated learning via multi-round consistency.
\newblock In \emph{Proceedings of the Computer Vision and Pattern Recognition Conference}, pages 15454--15463, 2025.

\bibitem[Xu et~al.(2022)Xu, Huang, Song, and Lan]{xu2022byzantine}
Jian Xu, Shao-Lun Huang, Linqi Song, and Tian Lan.
\newblock Byzantine-robust federated learning through collaborative malicious gradient filtering.
\newblock In \emph{2022 IEEE 42nd International Conference on Distributed Computing Systems (ICDCS)}, pages 1223--1235. IEEE, 2022.

\bibitem[Yin et~al.(2018)Yin, Chen, Kannan, and Bartlett]{trimmed_mean_median}
Dong Yin, Yudong Chen, Ramchandran Kannan, and Peter Bartlett.
\newblock Byzantine-robust distributed learning: Towards optimal statistical rates.
\newblock In \emph{International Conference on Machine Learning}, pages 5650--5659. PMLR, 2018.

\end{thebibliography}
}


\clearpage
\setcounter{page}{1}
\maketitlesupplementary


\appendix

\section{Design choices for XFED}
\label{xfed-design}
Prior model poisoning attacks~\cite{shejwalkar2021manipulating, shejwalkar2022back, baruch2019little, fang2020local} compute scaling factors by leveraging benign client updates. In our non-collusive threat model, an attacker cannot observe the individual benign updates, nor can it compute their maximum pairwise distance or exact distribution. The only information available to every compromised client is the sequence of global models broadcast by the server. Therefore, our goal is not to estimate the true maximum benign distance (which is impossible without collusion), but to infer a plausible acceptable deviation scale—that is, how far a client update can deviate from the global model without being rejected by the aggregation rule. We propose that the attacker estimate \( \mu \) using the sequence of global models \( (\theta^{(1)}, \theta^{(2)}, \ldots, \theta^{(t)}) \) received up to round \( t \).

\subsection{Why distances of global models represent the spread of local models}

As in round $t$ of FL training, all benign clients start from the same global model $\theta^{(t-1)}$ and perform local training to generate local models \( ^{(t)}\phi^{b}_1, ^{(t)}\phi^{b}_2, \ldots, ^{(t)}\phi^{b}_k \), we can write \(i^{\text{th}}\) benign local update as,

\[
^{(t)}\phi_i^{b} = \theta^{(t-1)} + \delta_i^{(t)}
\]

These local models are aggregated to create the global model $\theta^{(t)}$, 

\begin{align*}
\theta^{(t)} &= aggregation( ^{(t)}\phi^{b}_1, ^{(t)}\phi^{b}_2, \ldots, ^{(t)}\phi^{b}_k)\\
&= aggregation( \theta^{(t-1)} + \delta_1^{(t)}, \theta^{(t-1)} + \delta_2^{(t)}, \ldots,\\
&\qquad \theta^{(t-1)} + \delta_k^{(t)})
\end{align*}

Conceptually, the global update $\Delta \theta^{(t)} = \theta^{(t)} - \theta^{(t-1)}$ is a summary of how the benign clients collectively moved away from the previous global model $\theta^{(t-1)}$. More precisely, we can say $\Delta \theta^{(t)}$ is an approximation of the aggregations of the local deltas,

\[
\Delta \theta^{(t)}\approx \tilde{A}(\delta_1^{(t)}, \delta_2^{(t)}, \ldots,\delta_k^{(t)})
\]

where $\tilde{A}$ is an aggregation that is close to the main aggregation on the update vectors by the global server $aggregation$. If the main aggregation is average, $\tilde{A}$ is also average. Otherwise, for Byzantine-robust aggregations and defenses that filter out extremes, $\tilde{A}$ is an estimation of the most closely related local deltas. Thus, in all cases, the norm $\|\Delta \theta^{(t)}\|$ is a function of the distribution of benign local updates in that round: if benign updates are larger or more spread out, the global step tends to be larger; if they are small and concentrated, the global step is small. This makes the norm $\|\Delta \theta^{(t)}\|$ a natural proxy for the overall scale of benign local updates in that round. Geometrically, we can view the benign local updates in round $t$ as points scattered around $\theta^{(t-1)}$. Some of these points lie closer to $\theta^{(t-1)}$, others deviate more strongly depending on their local data and gradients. When these updates are aggregated, the new global model $\theta^{(t)}$ moves in the net direction favored by the majority of benign clients. The distance $\|\Delta \theta^{(t)}\|$ is therefore determined by the typical magnitude and orientation of benign updates: if benign clients collectively take larger steps, the global step is larger; if their updates are small and tightly concentrated, the global step is correspondingly small. We do not claim that $\Delta \theta^{(t)}$ equals the maximum pairwise distance between benign updates. However, it reliably reflects the scale of benign movement that the server “accepts” and amplifies in that round.

With this proxy in hand, we can choose the scaling coefficient $\mu$ so that the malicious update $\phi^{m}$ lies within the empirically observed range of benign behavior. Intuitively, if we ensure that $\|\phi^{m} - \phi^{b}\|$ is comparable to or slightly larger than the typical global step size inferred from recent rounds, the malicious update will fall inside the benign “ball” that robust aggregators are designed to accept. To make this estimate robust and adaptive, we do not rely on a single round. Instead, we maintain the global steps \(\mathcal{H}^{(t)} = \{\, \Delta\theta^{(1)}, \Delta\theta^{(2)}, \ldots, \Delta\theta^{(t)}\}\) and apply robust statistics, the median and median absolute deviation (MAD), to this sequence. The median captures the typical global step size over the window, while MAD captures its variability in a way that is resistant to occasional spikes (for example, early rounds, transient non-stationarity, or brief periods of strong attack activity).

 In this way, distances between consecutive global models act as a practical, non-collusive surrogate for the spread of benign local models: they summarize the magnitude of benign variation that actually passes through the aggregation filter, adapt to different datasets and architectures, and can be estimated solely from information available to each attacker.

\subsection{\textbf{Choice of scaling multiplier }$\lambda$}
\label{lambda-limit}

\subsubsection{Robust Outlier Detection via Median and Median Absolute Deviation (MAD)}

The median absolute deviation (MAD) is a robust statistical measure used to assess data variability while minimizing the influence of outliers. For a univariate sample $X=\{x_1, x_2, \ldots, x_k\}$, the median, 
$med = \operatorname{median}(X)$ 
provides a robust estimate of central tendency that is insensitive to extreme values. The \emph{median absolute deviation (MAD)} is then defined as,
\[
mad = 
  \operatorname{median}\!\left( \lvert x_i - med \rvert \right),
\]
Unlike the standard deviation, MAD has a breakdown point of 50\%, which means it remains stable even if up to half of the data points are extreme or adversarial. MAD is consistent with the standard deviation for normally distributed data, following the relationship \(\sigma \approx 1.4826 \times mad\) where \(\sigma\) is the standard deviation. In literature~\cite{iglewicz1993volume} to detect outliers in normal distribution, the following equation is used,
\begin{equation}
\label{eqn:outlier}
|\frac{x_i - med}{1.4826 \times mad}| > D
\end{equation}

If \(x_i\) satisfy the above equation, it is an outlier. Iglewicz and Hoaglin~\cite{iglewicz1993volume} recommended using $D=3.5$.

\subsubsection{Upper limit for $\lambda$}

Now, we want to calculate a maximum limit for $x_i$ so that it is not considered an outlier. To maximize $x_i$, we have to take $x_i > med \Rightarrow x_i - med > 0$. Therefore, from Equation~\ref{eqn:outlier} we find,

\begin{equation}
\label{eqn:xi-limit}
\begin{split}
x_i &\leq 1.4826 \times 3.5 \times mad + med \\
    &\leq 5.189 \times mad + med \\
\end{split}
\end{equation}

We extend the idea of Equation~\ref{eqn:xi-limit} to our calculation of $\mu$. Specifically, we take coordinate-wise median and MAD of the global delta list \(\mathcal{H}^{(t)}\) and calculate the $\mu$ as,
\[
\mu = \bigl\| med + \lambda \cdot mad \bigr\|_2
\]

where a practical highest limit for $\lambda$ is 5.189 according to Equation~\ref{eqn:xi-limit}. For a more complete limit, we look at other standard values of $D$ used in practice and find they are in the range for $D \in [2,4.5]$. In Equation~\ref{eqn:outlier} and Equation~\ref{eqn:xi-limit}, the value 1.4826 comes from the relationship between standard deviation $\sigma$ with $mad$ for the normal distribution; for other distributions, the range is between $[1.15,2.1]$ as shown in the Table~\ref{tab:const-dist}. This leaves us with the range for $\lambda$ as
\[
\begin{split}
    \lambda &\in [1.15 \times 2, 4.5 \times 2.1]\\
            &\in [2.3, 9.45]\\
            &\in [2, 10]
\end{split}
\]

\begin{table}[htb]
    \centering
    \caption{Constants for Common Distributions}
    \label{tab:const-dist}
    \begin{tabular}{@{}l||ll}
        \toprule
        Distribution & Relationship with standard deviation $\sigma$ \\
        \midrule
        Normal & $\sigma \approx 1.4826 \times mad$ \\
        Uniform & $\sigma \approx 1.1547 \times mad$ \\
        Exponential & $\sigma \approx 2.0781 \times mad$ \\
        Laplace & $\sigma \approx 2.0405 \times mad$ \\
        Logistic & $\sigma \approx 1.6205 \times mad$ \\
        \bottomrule
    \end{tabular}
\end{table}

We experimented with different values of \(\lambda\) in the range \([2, 10]\), and we report the results in Appendix~\ref{attack-parameter-appendix}.

\subsection{\textbf{Discussion on Perturbation Vector}}
\label{disc-pert}

After $\mu$, next important part of the XFED attack ($\phi^m = \phi^b + \mu \psi$) is to determine the attack direction $\psi$. Specifically, we need to choose an attack direction as a unit vector so we can multiply it by $\mu$ to push $\phi^b$ in a malicious direction. To calculate $\psi$, we use the current round's benign model update $\phi^b$. Specifically, we consider two perturbation vectors: the \textit{inverse sign vector} and the \textit{inverse unit vector}.
\[
\psi_{uv} = -\frac{\phi^{b}}{\|\phi^{b}\|_2}, 
\qquad
\psi_{sgn} = -\frac{\text{sign}(\phi^{b})}{\|\text{sign}(\phi^{b})\|_2}.
\]
Both $\psi_{\text{sgn}}$ and $\psi_{\text{uv}}$ are unit-norm direction vectors. Intuitively, these vectors are the reverse directions of $\phi^b$, and scaling them up to an acceptable limit should push the global model ``backwards'' along a direction that will hamper its accuracy. If enough adversarial clients apply such updates, the global model gradually starts to unlearn.

In our framework, the inverse unit vector ($X_{uv}$) has a clear advantage over the inverse sign vector ($X_{sgn}$). In $\psi_{\text{sgn}}$, each coordinate has the same magnitude, so the attack pushes equally in all dimensions, regardless of how important that dimension is for the current model. In contrast, $\psi_{\text{uv}}$ preserves the relative magnitudes of $\phi^b$: coordinates where $|\phi^b_j|$ is large contribute more to $\psi_{\text{uv}}$, while coordinates with small $|\phi^b_j|$ contribute less. Under benign training, large-magnitude coordinates typically correspond to parameters that have higher influence on the model’s predictions. By pushing more aggressively on these ``important'' dimensions and staying closer to $\phi^b$ on less relevant ones, $X_{uv}$ is more effective at degrading the global model while still preserving a benign-looking structure. Empirically, we observe that $X_{uv}$ consistently achieves stronger attack impact than $X_{sgn}$ across datasets, architectures, and aggregation rules.

\section{Additional Details of Experimental Setup}
\subsection{Datasets}\label{dataset}

We use multiple datasets from different domains in our evaluation, including four image classification benchmarks (MNIST~\cite{deng2012mnist}, EMNIST~\cite{cohen2017emnist}, Fashion-MNIST~\cite{xiao2017fashion}, CIFAR-10~\cite{cifar10}), one customer transaction behavior dataset (Purchase-100~\cite{wen2018customer}), and one human activity recognition dataset (HAR)~\cite{anguita2013public}. We vary both the number of clients and the model architectures across different datasets. We summarize the key configurations for each dataset and model in Table~\ref{tab:dataset_details}, including the number of clients, the total training and testing sample sizes, the number of training rounds, and the core hyperparameters (batch size and learning rate).

\begin{table*}[tb]
\centering
\caption{The default cross-silo FL system parameter settings.}
\label{tab:dataset_details}
\resizebox{\textwidth}{!}{%
\begin{tabular}{ccccccc}
\hline
\textbf{Parameter} & \textbf{MNIST} & \textbf{EMNIST} & \textbf{Purchase-100} & \textbf{Fashion-MNIST} & \textbf{CIFAR-10} & \textbf{HAR} \\ \hline
Architecture       & 4 layer DNN    & 3 layer DNN     & 3 layer DNN           & AlexNet, 3 layer DNN                & AlexNet           & LR           \\
Number of clients in FL            & 100   & 200    & 100    & 40    & 50    & 30    \\
\# local iterations                & 1     & 1      & 1      & 1     & 1     & 1     \\
\# global iterations               & 275   & 400    & 500    & 250   & 255   & 1000  \\
Batch size                         & 256   & 256    & 128    & 256   & 250   & 32    \\
Learning rate                      & 1     & 0.3    & 0.1    & 0.3   & 1     & 0.001 \\
Fraction of malicious clients (\%) & 20    & 20     & 20     & 20    & 20    & 20    \\
\# malicious clients               & 20    & 40     & 20     & 8     & 10    & 6     \\
Training Samples                   & 60000 & 697932 & 157324 & 60000 & 50000 &  7352     \\
Testing Samples                    & 10000 & 116323 & 39480  & 10000 & 10000 &  2947     \\
Training samples per client        & 600   & 3489   & 1578   & 1500  & 1000  & 201-299      \\ \hline
\end{tabular}%
}
\end{table*}

\paragraph{MNIST} MNIST\cite{deng2012mnist} is a benchmark dataset containing 70,000 grayscale images of handwritten digits (28 × 28 pixels) across 10 classes. We configure our federated learning framework by distributing the dataset among 100 clients, each with 600 samples. We conduct training over 275 rounds, utilizing an SGD optimizer. Our global model architecture is a fully connected network with layers \{784, 600, 100, 10\}, with ReLU activations in each layer.

\paragraph{EMNIST} EMNIST \cite{cohen2017emnist} is a 62-class classification dataset containing grayscale images (28 × 28) of handwritten characters, covering digits and both uppercase and lowercase letters. We use 200 clients, each with 600 training samples. During each communication round, we train locally on mini-batches of 256 images from each client’s dataset. For EMNIST, we implement a fully connected network (FC) with layer sizes \{784, 512, 512, 62\} as the global model architecture, with ReLU activations in the hidden layers.

\paragraph{Purchase-100} Purchase-100 \cite{wen2018customer} is a 100-class classification dataset containing binary feature vectors of length 600, representing customer purchase patterns. We configure our federated learning framework with 100 clients, each with 1,578 samples. We train locally on mini-batches of 128 samples per client during each communication round. Training is conducted over 500 rounds using an SGD optimizer. Our global model architecture is a three-layer fully connected network with layer sizes \{600, 1024, 100\}.

\paragraph{Fashion-MNIST} The Fashion-MNIST dataset \cite{xiao2017fashion} consists of 60,000 training and 10,000 test samples across 10 fashion classes. Each sample is a 28×28 grayscale image with 784 features representing the relative pixel intensities on a scale of [0, 255]. The setup configuration consists of 40 clients, each with 1,500 samples. Training is conducted over 250 rounds, with local training on mini-batches of 256 samples per client. We use AlexNet and a three-layer fully connected network architecture as the global models.

\paragraph{CIFAR-10} CIFAR-10 \cite{cifar10} is a 10-class classification dataset with 60,000 color images (32 × 32 pixels). We configure our federated learning setup with 50 clients, each with 1,000 samples. Training is conducted over 255 rounds, with local training on mini-batches of 256 samples per client. We use an AlexNet architecture as the global model.

\paragraph{Human Activity Recognition (HAR)} HAR \cite{anguita2013public} consists of sensor-based activity data collected from the smartphones of 30 real-world participants. Each sample contains 561 features, including accelerometer and gyroscope measurements, and is labeled with one of six activities: WALKING, WALKING\_UPSTAIRS, WALKING\_DOWNSTAIRS, SITTING, STANDING, or LAYING. The dataset comprises 10,299 examples. In our federated setting, each participant is treated as a distinct client, resulting in 30 clients in total. Unlike previous datasets, we do not need to distribute the data to clients in this dataset, since each user is naturally a client. We use 75\% of each client's local data for training and the remaining 25\% for testing. The global model is a single-layer logistic regression classifier that maps the 561-dimensional input to the six activity classes. Training is performed over 1000 communication rounds using stochastic gradient descent (SGD) with a local batch size of 32.

\subsection{State-of-the-art Model Poisoning Attacks}\label{baseline-attacks-appendix}

\paragraph{A Little Is Enough (LIE)} LIE~\cite{baruch2019little} demonstrates that adding a small, carefully aligned perturbation to the local update can significantly degrade the global model while remaining undetected. To evade Byzantine-robust aggregation, LIE constructs a poisoned update by staying within the statistical range of benign client updates. In communication round \(t\), let \(\phi_i^{(t)}\) denote the benign update of client \(i\). For each parameter \(j\), the attacker computes the empirical mean \(\mu_j^{(t)}\) and standard deviation \(\sigma_j^{(t)}\) across all benign updates (which LIE assumes the attacker can observe). The malicious update is then crafted so that each parameter lies within:
\[
\phi^{m,(t)}_j = \mu_j^{(t)} \pm z_{\max} \, \sigma_j^{(t)},
\]
where \(z_{\max} \in [0,1]\) is a scaling factor that determines how close the update is pushed toward the edge of the benign distribution. The final poisoned update \(\phi^{m,(t)}\) is shared across all adversarial clients, ensuring coordinated behavior in each round.

\paragraph{Fang} When we refer to Fang, we mean the Fang-Krum version of the original paper when the aggregation method is Multi-Krum, and the Fang-TrimmedMean version when the aggregation methods are Trimmed-Mean, Median, and Fed-Avg. Below, we briefly outline the Fang-Krum and Fang-TrimmedMean attacks. We only run the Fang attack against FedAvg, Tr-Mean, Median, and M-Krum aggregations, as the Fang attack has no versions for other aggregations or defenses.

\paragraph{\textbf{Fang-Krum}} The Fang-Krum attack~\cite{fang2020local} targets the Krum aggregation rule by constructing a set of \(c\) adversarial updates, 
\(\{\phi_1^{\text{adv}}, \phi_2^{\text{adv}}, \dots, \phi_c^{\text{adv}}\}\), 
such that Krum selects one of them as the aggregated global update. The attacker’s goal is to force Krum to output an adversarial update that deviates substantially from the benign aggregation direction. For each compromised client \(i\), the attacker starts from the client’s benign update \(\phi_i^{\text{real}}\) and perturbs it along a malicious direction \(s\), scaled by a factor \(\lambda\):
\[
\phi_i^{\text{adv}} = \phi_i^{\text{real}} - \lambda s.
\]
The attacker then searches for a value of \(\lambda\) that ensures at least one adversarial update is selected by the Krum rule. After identifying such an update, the remaining \(c - 1\) adversarial updates are constructed to lie close to the selected one, minimizing their pairwise distances.

\paragraph{\textbf{Fang-TrimmedMean}} To craft a Fang--TrimmedMean attack~\cite{fang2020local}, the adversary exploits statistical information computed from the model updates of the compromised clients. For each parameter index \( j \), the attacker first computes the empirical mean \( \mu_j \) and standard deviation \( \sigma_j \) over the set of adversarial updates 
\(\{\phi_1^{\text{adv}}, \phi_2^{\text{adv}}, \dots, \phi_c^{\text{adv}}\}\). To bias the aggregation outcome in a desired direction, the attacker injects extreme yet internally consistent values for each parameter. If the attack direction is negative (\(-1\)), each compromised client samples the \( j \)-th coordinate from the interval 
\[
[\mu_j + 3\sigma_j,\; \mu_j + 4\sigma_j].
\]
Conversely, to push the model in the positive direction (\(+1\)), the value is sampled from 
\[
[\mu_j - 4\sigma_j,\; \mu_j - 3\sigma_j].
\]

\paragraph{\textbf{Min-Max}} The Min-Max attack~\cite{shejwalkar2021manipulating} is an aggregation-agnostic model poisoning attack designed to make adversarial updates indistinguishable from benign ones under distance-based aggregation rules. Let 
\(\mathcal{B} = \{\phi_1^{\text{benign}}, \dots, \phi_n^{\text{benign}}\}\) 
denote the set of benign updates and 
\(\mathcal{A} = \{\phi_1^{\text{adv}}, \dots, \phi_c^{\text{adv}}\}\) 
the set of adversarial updates. The attacker seeks to minimize the maximum distance between any adversarial update and the benign set while staying within the benign update radius. Formally, the attack enforces:
\[
\max_{\phi_i^{\text{adv}} \in \mathcal{A},\; \phi_j^{\text{benign}} \in \mathcal{B}}
\big\|\phi_i^{\text{adv}} - \phi_j^{\text{benign}}\big\|
\;\le\;
\max_{\phi_k, \phi_l \in \mathcal{B}}
\big\|\phi_k - \phi_l\big\|.
\]

\paragraph{Min-Sum} The Min-Sum attack~\cite{shejwalkar2021manipulating} is an aggregation-agnostic model poisoning attack that makes malicious updates indistinguishable from benign ones under a \emph{sum-of-squared-distances} criterion. Let $\mathcal{B}=\{\phi^{\text{benign}}_1,\dots,\phi^{\text{benign}}_n\}$ denote benign updates and $\mathcal{A}=\{\phi^{\text{adv}}_1,\dots,\phi^{\text{adv}}_c\}$ denote adversarial updates. The attacker sets all adversarial updates identical for maximum impact and constrains each malicious update to satisfy

$$
\begin{aligned}
\sum_{j=1}^{n}\bigl\|\phi^{\text{adv}}-\phi^{\text{benign}}_{j}\bigr\|_2^{2}
&\le
\max_{i\in[n]}\sum_{j=1}^{n}\bigl\|\phi^{\text{benign}}_{i}-\phi^{\text{benign}}_{j}\bigr\|_2^{2},\\[4pt]
\phi^{\text{adv}}&=\phi^{b}+\mu\,\psi.
\end{aligned}
$$

where $\phi^{b}$ is a benign reference aggregate, $\psi$ is a fixed perturbation direction, and $\mu$ is a scaling coefficient optimized to maximize attack impact under the constraint.   

\paragraph{\textbf{MPAF}} The Model Poisoning Attack based on Fake Clients (MPAF)~\cite{cao2022mpaf} is an untargeted model-poisoning attack in which the adversary injects attacker-controlled fake clients into the FL system. MPAF operates under minimal knowledge, requiring only access to the server-broadcast global models. The attacker first selects a low-accuracy base model \(w'\) sharing the same architecture as the global model. Then, in each communication round \(t\), every fake client submits an update that pulls the global model toward \(w'\):
\[
g_i^{t} = \lambda\,(w' - w_t), \qquad \lambda > 0,
\]
thereby steering the training trajectory to reduce the distance \(\|w_T - w'\|\) over time. By repeatedly applying this directional shift, MPAF gradually degrades the global model's performance throughout the FL process.

\paragraph{\textbf{PoisonedFL}} PoisonedFL ~\cite{xie2024model} is a model-poisoning attack that improves adversarial effectiveness by enforcing multi-round consistency among malicious updates. Its core insight is that prior attacks primarily rely on within-round coordination, leading their malicious effects to partially cancel out across training rounds. PoisonedFL overcomes this limitation by injecting attacker-controlled fake clients and generating adversarial updates that remain directionally aligned over the entire training trajectory. The attack operates with minimal knowledge, requiring only access to server-broadcast global models and no information about benign clients’ data or updates. 

\subsection{Baseline Aggregation Rules}\label{aggregation-appendix}

\paragraph{FedAvg} In Fed-Avg, the central server aggregates the local models \( \{\phi_1^{(t)}, \phi_2^{(t)}, \dots, \phi_k^{(t)}\} \) from \( k \) participating edge devices during each communication round \( t \) to update the global model \( \theta^{(t+1)} \). The global model \( \theta^{(t+1)} \) is computed as a weighted average of the local models:
\(
\theta^{(t+1)} = \frac{1}{k} \sum_{i=1}^{k} \phi_i^{(t)}.
\)

\paragraph{Median} In Median aggregation~\cite{trimmed_mean_median}, for each \( j \)-th parameter of all \( k \) local models, the central server collects the \( j \)-th parameters (\(\phi_{1j}, \phi_{2j}, \dots, \phi_{kj}\)) of all local model updates. The server then computes the median of these \( j \)-th parameters across the \( k \) local models. The resulting median value for each parameter position \( j \) is used to form the \( j \)-th parameter of the global model in the next communication round:
\(
\theta_j^{(t+1)} = \text{median}(\phi_{1j}^{(t)}, \phi_{2j}^{(t)}, \dots, \phi_{kj}^{(t)}),
\)
where \( \theta_j^{(t+1)} \) denotes the \( j \)-th parameter of the global model after aggregation in round \( t+1 \), and the median is computed over the \( j \)-th parameters of all local model updates.

\paragraph{Trimmed-Mean} In Trimmed-Mean~\cite{trimmed_mean_median} aggregation, during each communication round \( t \), for each parameter \( j \) of the model, the central server collects the \( j \)-th parameters from all \( k \) local model updates \( \{\phi_{1j}^{(t)}, \phi_{2j}^{(t)}, \dots, \phi_{kj}^{(t)}\} \). The server then sorts these \( j \)-th parameter values and removes the largest and smallest \( c \) values, where \( c \) is the assumed number of compromised edge devices. The global model's \( j \)-th parameter is then computed by averaging the remaining \( k - 2c \) values:
\(
\theta_j^{(t+1)} = \frac{1}{k - 2c} \sum_{i=c+1}^{k - c} \phi_{ij}^{(t)},
\)
This process is repeated for each parameter \( j \) of the model, resulting in the updated global model \( \theta^{(t+1)} \).

\paragraph{Multi-Krum} Multi-Krum~\cite{krum} aggregation selects \( f \) local models from the set of local updates \(\{\phi_1^{(t)}, \phi_2^{(t)}, \dots, \phi_k^{(t)}\}\) during each communication round \( t \) and forms the global model \( \theta^{(t+1)} \) by averaging these selected local models:
\(
\theta^{(t+1)} = \frac{1}{f} \sum_{i=1}^{f} \phi_{i}^{(t)}.
\)
To select the \( f \) local model updates, the central server assumes that at most \( c \) edge devices are compromised. For each local model \( \phi_i^{(t)} \) (where \( i = 1, 2, \dots, k \)), the server computes the distances to all other local models. It then identifies the \( k - c - 2 \) local models closest to \( \phi_i^{(t)} \) and computes the sum of distances between \( \phi_i^{(t)} \) and these \( k - c - 2 \) nearest local models. Finally, the server selects the \( f \) local models with the smallest sum of distances to their nearest \( k - c - 2 \) neighbors.

\paragraph{Clipped-Clustering} In Clipped-Clustering~\cite{li2023experimental}, the server initially sets a clipping threshold hyper-parameter \( \tau \). During each communication round \( t \), for each local model update \( \phi_i^{(t)} \) (where \( i = 1, 2, \dots, k \)), the server applies a ``clip by norm" technique. Specifically, if the norm of \( \phi_i^{(t)} \) exceeds \( \tau \), the parameters of \( \phi_i^{(t)} \) are scaled down to meet this threshold. If the norm of \( \phi_i^{(t)} \) is less than \( \tau \), the update remains unchanged. After clipping, the server performs \( k \)-means clustering on the set of clipped updates \( \{\phi_1^{(t)}, \phi_2^{(t)}, \dots, \phi_k^{(t)}\} \) and divides the updates into two clusters. The server then identifies the larger cluster and computes the average of the updates in this cluster to determine the global model \( \theta^{(t+1)} \).

\paragraph{SignGuard} SignGuard~\cite{xu2022byzantine} filters malicious updates by analyzing the element-wise sign patterns of model gradients. In each communication round \( t \), the central server receives local updates \(\{\phi_1^{(t)}, \phi_2^{(t)}, \dots, \phi_k^{(t)}\}\) and computes the proportion of positive, negative, and zero elements in each update, creating a feature set. The server then performs clustering on these features to identify updates with similar sign patterns. Updates that align with the majority cluster are retained and averaged to form the global model.

\subsection{State-of-the-art Defenses}\label{defenses-appendix}
\paragraph{FLAME} FLAME~\cite{nguyen2022flame} mitigates the impact of malicious updates by adaptively filtering out outliers during aggregation. At each communication round, FLAME measures the distance between every client update and the current global model. Client updates whose distances exceed an adaptively computed threshold are eliminated before aggregation.

\paragraph{FreqFed}
FreqFed~\cite{fereidoonifreqfed} detects adversarial behavior by analyzing the frequency patterns of parameter updates across clients. FreqFed tracks how often each client modifies individual model parameters over multiple rounds and identifies clients whose update frequencies deviate from the dominant behavior observed among benign participants. Clients exhibiting unusually rare, inconsistent, or infrequent parameter changes are flagged as suspicious and excluded from aggregation.

\paragraph{Foolsgold} Foolsgold~\cite{fung2018mitigating} detects and suppresses sybil-based poisoning attacks by leveraging the similarity of client updates. The key insight is that malicious clients often produce highly correlated model updates, whereas benign clients tend to contribute diverse gradients. Foolsgold computes pairwise cosine similarities between client updates across rounds and down-weights clients whose updates exhibit unusually high similarity to others. By adaptively reducing the aggregation weight of correlated clients, Foolsgold aims to limit the collective influence of coordinated adversaries.

\paragraph{FLTrust} In FLTrust~\cite{cao2021fltrust}, a trusted server maintains a small, clean ``root dataset," which is used to compute a reference gradient at each communication round. Client updates are then compared to this trusted gradient to assess their similarity. Each client is assigned a trust score based on how closely its updates align with the root gradient and its historical behavior. These trust scores are used to weigh client updates during aggregation, giving greater influence to trustworthy clients while reducing the impact of potentially malicious ones. For our experiments, we constructed the ``root dataset" by randomly sampling 100 instances from the training data. We set the bias parameter to 0.1, ensuring that each class contributes an equal number of samples and that no class comprises more than 10\% of the root dataset.

\section{Additional Experimental Results and Analysis}

\subsection{Why XFED Outperforms Existing Attacks}
\label{effectiveness_xfed}

A key finding from our experimental evaluation is that XFED consistently outperforms existing model poisoning attacks, despite operating under a significantly weaker and more realistic non-collusive threat model. At first glance, this is counterintuitive, since XFED has access to less information and does not rely on any inter-client coordination. Its superiority primarily stems from the use of robust, data-driven statistical bounds to constrain the attack magnitude \((\mu)\). By computing the median and the median absolute deviation (MAD) of recent global updates, XFED estimates a non-outlier region in parameter space and scales its perturbation to lie precisely near this boundary (Section~\ref{lambda-limit}).

In contrast, existing attacks such as Fang, Min-Sum, and Min-Max typically require iterative search procedures (e.g., binary search) to determine the largest feasible perturbation that avoids rejection by the aggregation rule. However, determining the largest feasible perturbation by iterative search procedures can be unstable across rounds or aggregation rules. Fang’s aggregation-targeted attack sometimes benefits from privileged knowledge of both the aggregation rule and the benign updates of other malicious clients, which is unrealistic in practical deployments. Although this additional knowledge can occasionally make Fang competitive with XFED, its reliance on a uniform sign-based perturbation direction (\( \psi_{sgn} \)) limits its effectiveness. PoisonedFL~\cite{xie2024model} achieves strong performance in cross-device settings but is highly sensitive to parameter choices and does not generalize well to cross-silo scenarios.

\subsection{Analyzing the Effectiveness of XFED} \label{outperform_xfed}

XFED has a significant impact on all types of aggregations and defenses. One interesting observation is that XFED performs better against defenses than it does against Byzantine-robust aggregations. This is primarily because defenses employ outlier detection techniques to filter out harmful model updates before they are aggregated. XFED is specifically designed to avoid being identified as an outlier. As a result, even if all attackers submit different harmful model updates, most of these updates go undetected as outliers and are still included in the global model calculation. In contrast, Byzantine-robust aggregations typically use dimension-wise clipping. In non-collusive attacks, some attack dimensions may be clipped. Nevertheless, XFED continues to have a significant impact on Byzantine-robust aggregations, as the non-filtered dimensions of the poisoned model updates can cause substantial damage. We provide a detailed discussion on the choice of the upper bound for \(\lambda\) in Section~\ref{lambda-limit}.

\subsection{Impact of cross-device setting.} \label{cross-device-appendix}

\begin{table}[!tb]
\centering
\caption{Attack impact \( I_\theta \) on the global model under different attacks (20\% malicious clients), aggregation rules, and defenses for the Purchase dataset in cross-device settings. In each row, we highlight in bold the highest attack impact or the closest to the highest.}
\label{tab:crossdevice-appendix-table}
\resizebox{\linewidth}{!}{%
\begin{tabular}{@{}l||cccccccccc@{}}
\toprule
& \( A_\theta \) & \(X_{sgn}\) & \(X_{uv}\) & Min-Sum & Min-Max & Fang & LIE & MPAF & PoisonedFL\\ \midrule
\multicolumn{1}{c||}{} & \multicolumn{9}{c}{\textbf{Purchase(Cross-device, 10000 FL clients, 500 global iterations \& 3 layer DNN model)}} \\ \midrule

FedAvg    & 75.07 & 62.23          & \textbf{70.72} & 25.18         & 38.23          & \textbf{72.37} & 9.26          & \textbf{72.22} & \textbf{72.08} \\
Median    & 72.39 & 41.96          & \textbf{44.7}  & 38.89         & \textbf{41.93} & 40.22          & 19.32         & 22.82          & \textbf{43.22} \\
Tr-Mean   & 74.04 & \textbf{60.02} & \textbf{60.97} & 36.94         & 44.21          & 23.56          & 15.63         & 59.22          & \textbf{62.22} \\
M-Krum    & 74.79 & 9.09           & \textbf{40.22} & 35.97         & \textbf{42.47} & 32.88          & 13.3          & \textbf{45.22} & \textbf{41.37} \\
CC        & 74.94 & 13.68          & \textbf{70.79} & 39.52         & 52.77          & -              & 4.31          & 3.53           & 4.39           \\
SgnG      & 74.68 & 9.21           & \textbf{72.82} & 37.19         & 55.13          & -              & 3.45          & 4.65           & 4.57           \\
FLTrust   & 74.22 & 3.03           & 2.86           & 4.74          & \textbf{5.88}  & -              & \textbf{6.67} & 4.81           & 3.6            \\
FreqFed   & 73.89 & 7.35           & \textbf{71.22} & 45.62         & 57.23          & -              & 3.37          & 4.21           & 5.9            \\
FoolsGold & 74.41 & 5.22           & \textbf{10.56} & \textbf{9.25} & \textbf{10.22} & -              & 3.35          & 4.56           & \textbf{9.22}  \\
FLAME     & 74.28 & 4.2            & \textbf{70.96} & 48.82         & 60.56          & -              & 26.14         & 4.05           & 4.37

\\ \bottomrule
\end{tabular}%
}
\end{table}

We assess the impact of our proposed attacks in a cross-device setting using the Purchase dataset, which includes 20\% malicious clients. To simulate this setting, we consider 10,000 clients, with 1\% randomly selected in each round of federated learning (FL) training. The results are presented in Table~\ref{tab:crossdevice-appendix-table}, where the \(A_\theta\) column corresponds to the ``No Attack'' baseline, and the remaining columns report the attack impact \(I_\theta\) for the different attacks.
 Similar to the cross-silo setting, our proposed non-collusive attacks, particularly \(X_{uv}\), outperform all state-of-the-art model poisoning attacks across most combinations of aggregation rules and defenses in the cross-device setting. For example, for the Clipped Clustering aggregation, \(X_{uv}\) achieves up to 20x higher impact compared to LIE, MPAF, and PoisonedFL. Similarly, against the FreqFed defense, \(X_{uv}\) achieves 1.56× and 1.24× greater impact than the Min-Sum and Min-Max attacks, respectively. Interestingly, some results in the cross-device setting vary from the trends observed in the cross-silo setting. This is mainly due to the stochastic nature of client selection in cross-device FL: in each round, only a small fraction of clients participate, leading to fluctuating numbers of adversarial clients. Consequently, some rounds may contain only a few, or even no, attackers.

\subsection{Impact of the percentage of malicious clients} \label{purchase-all-attacker}


\pgfplotstableread[col sep=comma, header=true]{Purchase-Krum.csv}\PurchaseKrum

\pgfplotstableread[col sep=comma, header=true]{Purchase-TRMEAN.csv}\PurchaseTrmean

\pgfplotstableread[col sep=comma, header=true]{Purchase-CC.csv}\PurchaseCC

\pgfplotstableread[col sep=comma, header=true]{Purchase-SignGuard.csv}\PurchaseSign

\pgfplotstableread[col sep=comma, header=true]{Purchase-FoolsGold.csv}\PurchaseGold

\pgfplotstableread[col sep=comma, header=true]{Purchase-FreqFred.csv}\PurchaseFreq

\pgfplotstableread[col sep=comma, header=true]{Purchase-FLAME.csv}\PurchaseFLAME

\pgfplotstableread[col sep=comma, header=true]{Purchase-FLTrust.csv}\PurchaseFLTrust

\begin{figure*}[tb]
\centering

\setlength\panelsep{2mm}

\begin{tikzpicture}
\begin{groupplot}[
  group style={group size=8 by 1, horizontal sep=\panelsep, vertical sep=0pt},
  width=\dimexpr(\textwidth-9\panelsep)/8\relax,
  height=2.1cm,                 
  xtick={0,5,10,15,20,25,30},
  xticklabels={0,5,10,15,20,25,30},
  ytick style={draw=none},
  ymajorgrids=false,
  title style={yshift=-1mm},
  xmin=0,xmax=30,
  ymin=0,ymax=90, ytick={0,10,...,90}
]
\hspace{-5mm}
\nextgroupplot[
  title={\small M-Krum},
  ylabel={\scriptsize Attack Impact},
  trim axis left,
]
\addplot [color=green, mark=ball, mark size=0.5pt] table [y=VIRAT_min_max] {\PurchaseKrum};
\addplot [color=black, mark=ball, mark size=0.5pt] table [y=FANG] {\PurchaseKrum};
\addplot [color=orange, mark=ball, mark size=0.5pt] table [y=LIE] {\PurchaseKrum};
\addplot [color=purple, mark=ball, mark size=0.5pt] table [y=minsum] {\PurchaseKrum};
\addplot [color=brown, mark=ball, mark size=0.5pt] table [y=mpaf] {\PurchaseKrum};
\addplot [color=teal, mark=otimes, mark size=0.5pt] table [y=PoisonedFL]{\PurchaseKrum};
\addplot [color=red, mark=square*, mark size=0.5pt] table [y=XFED_sign] {\PurchaseKrum};
\addplot [color=blue, mark=ball, mark size=0.5pt] table [y=XFED_unit_vec] {\PurchaseKrum};

\nextgroupplot[title={\small Tr-Mean}, yticklabels=\empty, ylabel={}]
\addplot [color=green, mark=ball, mark size=0.5pt] table [y=VIRAT_min_max] {\PurchaseTrmean};
\addplot [color=black, mark=ball, mark size=0.5pt] table [y=FANG] {\PurchaseTrmean};
\addplot [color=orange, mark=ball, mark size=0.5pt] table [y=LIE] {\PurchaseTrmean};
\addplot [color=purple, mark=ball, mark size=0.5pt] table [y=minsum] {\PurchaseTrmean};
\addplot [color=brown, mark=ball, mark size=0.5pt] table [y=mpaf] {\PurchaseTrmean};
\addplot [color=teal, mark=otimes, mark size=0.5pt] table [y=PoisonedFL]{\PurchaseTrmean};
\addplot [color=red, mark=square*, mark size=0.5pt] table [y=XFED_sign] {\PurchaseTrmean};
\addplot [color=blue, mark=ball, mark size=0.5pt] table [y=XFED_unit_vec] {\PurchaseTrmean};

\nextgroupplot[title={\small CC}, yticklabels=\empty, ylabel={}]
\addplot [color=green, mark=ball, mark size=0.5pt] table [y=VIRAT_min_max] {\PurchaseCC};
\addplot [color=black, mark=ball, mark size=0.5pt] table [y=FANG] {\PurchaseCC};
\addplot [color=orange, mark=ball, mark size=0.5pt] table [y=LIE] {\PurchaseCC};
\addplot [color=purple, mark=ball, mark size=0.5pt] table [y=minsum] {\PurchaseCC};
\addplot [color=brown, mark=ball, mark size=0.5pt] table [y=mpaf] {\PurchaseCC};
\addplot [color=teal, mark=otimes, mark size=0.5pt] table [y=PoisonedFL]{\PurchaseCC};
\addplot [color=red, mark=square*, mark size=0.5pt] table [y=XFED_sign] {\PurchaseCC};
\addplot [color=blue, mark=ball, mark size=0.5pt] table [y=XFED_unit_vec] {\PurchaseCC};

\nextgroupplot[title={\small SgnG}, yticklabels=\empty, ylabel={}]
\addplot [color=green, mark=ball, mark size=0.5pt] table [y=VIRAT_min_max] {\PurchaseSign};
\addplot [color=black, mark=ball, mark size=0.5pt] table [y=FANG] {\PurchaseSign};
\addplot [color=orange, mark=ball, mark size=0.5pt] table [y=LIE] {\PurchaseSign};
\addplot [color=purple, mark=ball, mark size=0.5pt] table [y=minsum] {\PurchaseSign};
\addplot [color=brown, mark=ball, mark size=0.5pt] table [y=mpaf] {\PurchaseSign};
\addplot [color=teal, mark=otimes, mark size=0.5pt] table [y=PoisonedFL]{\PurchaseSign};
\addplot [color=red, mark=square*, mark size=0.5pt] table [y=XFED_sign] {\PurchaseSign};
\addplot [color=blue, mark=ball, mark size=0.5pt] table [y=XFED_unit_vec] {\PurchaseSign};

\nextgroupplot[title={\small FoolsGold}, yticklabels=\empty, ylabel={}]
\addplot [color=green, mark=ball, mark size=0.5pt] table [y=VIRAT_min_max] {\PurchaseGold};
\addplot [color=orange, mark=ball, mark size=0.5pt] table [y=LIE] {\PurchaseGold};
\addplot [color=purple, mark=ball, mark size=0.5pt] table [y=minsum] {\PurchaseGold};
\addplot [color=brown, mark=ball, mark size=0.5pt] table [y=mpaf] {\PurchaseGold};
\addplot [color=teal, mark=otimes, mark size=0.5pt] table [y=PoisonedFL]{\PurchaseGold};
\addplot [color=blue, mark=ball, mark size=0.5pt] table [y=XFED_unit_vec] {\PurchaseGold};
\addplot [color=red, mark=square*, mark size=0.5pt] table [y=XFED_sign] {\PurchaseGold};

\nextgroupplot[title={\small FLAME}, yticklabels=\empty, ylabel={}]
\addplot [color=green, mark=ball, mark size=0.5pt] table [y=VIRAT_min_max] {\PurchaseFLAME};
\addplot [color=black, mark=ball, mark size=0.5pt] table [y=FANG] {\PurchaseFLAME};
\addplot [color=orange, mark=ball, mark size=0.5pt] table [y=LIE] {\PurchaseFLAME};
\addplot [color=purple, mark=ball, mark size=0.5pt] table [y=minsum] {\PurchaseFLAME};
\addplot [color=brown, mark=ball, mark size=0.5pt] table [y=mpaf] {\PurchaseFLAME};
\addplot [color=teal, mark=otimes, mark size=0.5pt] table [y=PoisonedFL]{\PurchaseFLAME};
\addplot [color=red, mark=square*, mark size=0.5pt] table [y=XFED_sign] {\PurchaseFLAME};
\addplot [color=blue, mark=ball, mark size=0.5pt] table [y=XFED_unit_vec] {\PurchaseFLAME};

\nextgroupplot[title={\small FreqFed}, yticklabels=\empty, ylabel={}]
\addplot [color=green, mark=ball, mark size=0.5pt] table [y=VIRAT_min_max] {\PurchaseFreq};
\addplot [color=black, mark=ball, mark size=0.5pt] table [y=FANG] {\PurchaseFreq};
\addplot [color=orange, mark=ball, mark size=0.5pt] table [y=LIE] {\PurchaseFreq};
\addplot [color=purple, mark=ball, mark size=0.5pt] table [y=minsum] {\PurchaseFreq};
\addplot [color=teal, mark=otimes, mark size=0.5pt] table [y=PoisonedFL] {\PurchaseFreq};
\addplot [color=brown, mark=ball, mark size=0.5pt] table [y=mpaf] {\PurchaseFreq};
\addplot [color=red, mark=square*, mark size=0.5pt] table [y=XFED_sign] {\PurchaseFreq};
\addplot [color=blue, mark=ball, mark size=0.5pt] table [y=XFED_unit_vec] {\PurchaseFreq};

\nextgroupplot[title={\small FLTrust}, yticklabels=\empty, ylabel={}, trim axis right]
\addplot [color=green, mark=ball, mark size=0.5pt] table [y=VIRAT_min_max] {\PurchaseFLTrust};
\addplot [color=orange, mark=ball, mark size=0.5pt] table [y=LIE] {\PurchaseFLTrust};
\addplot [color=purple, mark=ball, mark size=0.5pt] table [y=minsum] {\PurchaseFLTrust};
\addplot [color=brown, mark=ball, mark size=0.5pt] table [y=mpaf] {\PurchaseFLTrust};
\addplot [color=teal, mark=otimes, mark size=0.5pt] table [y=PoisonedFL] {\PurchaseFLTrust};
\addplot [color=red, mark=square*, mark size=0.5pt] table [y=XFED_sign] {\PurchaseFLTrust};
\addplot [color=blue, mark=ball, mark size=0.5pt] table [y=XFED_unit_vec] {\PurchaseFLTrust};

\end{groupplot}
\end{tikzpicture}

    \par\vspace{2mm}\centering
      \begin{tikzpicture}
        \begin{axis}[
            xmin=0, xmax=45,
            ymin=0, ymax=1,
            legend columns=8,
            axis lines=none,
            width=0.35\textwidth,
            height=3cm,
            clip=false,
            legend style={font=\scriptsize}, 
        ]
          \addlegendimage{color=blue, mark=ball, mark size=0.5pt}
            \addlegendentry{ \( X_{uv} \)}
          \addlegendimage{color=red, mark=square*, mark size=0.5pt}
            \addlegendentry{ \( X_{sgn} \)}
          \addlegendimage{color=green, mark=ball, mark size=0.5pt}
            \addlegendentry{\( Min-Max \)}
          \addlegendimage{color=black, mark=ball, mark size=0.5pt}
            \addlegendentry{\( Fang \)}
        \addlegendimage{color=orange, mark=ball, mark size=0.5pt}
            \addlegendentry{\( LIE\)}
        \addlegendimage{color=purple, mark=ball, mark size=0.5pt}
            \addlegendentry{\( Min-Sum\)}
             \addlegendimage{color=brown, mark=ball, mark size=0.5pt}
            \addlegendentry{\( MPAF\)}
             \addlegendimage{color=teal, mark=otimes, mark size=0.5pt}
            \addlegendentry{\( PoisonedFL\)}
        \end{axis}
      \end{tikzpicture}%

\caption{Attack Impact \( I_\theta \) on global models with increasing \% of malicious clients for different aggregations, defenses, and attacks in a cross-silo setting (Purchase dataset).}
\label{impact-Purchase}
\end{figure*}
\pgfplotstableread[col sep=comma, header=true]{Purchase-Mean.csv}\PurchaseMean

\pgfplotstableread[col sep=comma, header=true]{Purchase-MEDIAN.csv}\PurchaseMedian

\pgfplotstableread[col sep=comma, header=true]{MNIST-Mean.csv}\MnistMean

\pgfplotstableread[col sep=comma, header=true]{MNIST-MEDIAN.csv}\MnistMedian


\pgfplotstableread[col sep=comma, header=true]{Purchase-crossdevice-MEAN.csv}\PurchaseCrossDeviceMean

\pgfplotstableread[col sep=comma, header=true]{Purchase-crossdevice-Median.csv}\PurchaseCrossDeviceMedian

\begin{figure}[tb]
\centering

\setlength\panelsep{2mm}

\begin{tikzpicture}
\begin{groupplot}[
  group style={group size=3 by 2,
    horizontal sep=10pt,
    vertical sep=35pt,},
    width=0.28\linewidth,   
  height=0.25\linewidth,     
  xtick={0,5,10,15,20,25,30},
  xticklabels={0,5,10,15,20,25,30},
  ytick style={draw=none},
  ymajorgrids=false,
  title style={yshift=-1mm},
  xmin=0,xmax=30,
  ymin=0,ymax=90, ytick={0,10,...,90}
]
\hspace{-5mm}
\nextgroupplot[
  title={MNIST, FedAvg,\\Cross-silo},
  title style={font=\small, align=center},
  ylabel={\scriptsize Attack Impact},
  trim axis left,
]
\addplot [color=green, mark=ball, mark size=0.5pt] table [y=VIRAT_min_max] {\MnistMean};
\addplot [color=black, mark=ball, mark size=0.5pt] table [y=FANG] {\MnistMean};
\addplot [color=orange, mark=ball, mark size=0.5pt] table [y=LIE] {\MnistMean};
\addplot [color=purple, mark=ball, mark size=0.5pt] table [y=minsum] {\MnistMean};
\addplot [color=brown, mark=ball, mark size=0.5pt] table [y=mpaf] {\MnistMean};
\addplot [color=teal, mark=otimes, mark size=0.5pt] table [y=PoisonedFL]{\MnistMean};
\addplot [color=red, mark=square*, mark size=0.5pt] table [y=XFED_sign] {\MnistMean};
\addplot [color=blue, mark=ball, mark size=0.5pt] table [y=XFED_unit_vec] {\MnistMean};

\nextgroupplot[
  title={MNIST, Median,\\Cross-silo},
  title style={font=\small, align=center}, yticklabels=\empty, ylabel={}]
\addplot [color=green, mark=ball, mark size=0.5pt] table [y=VIRAT_min_max] {\MnistMedian};
\addplot [color=black, mark=ball, mark size=0.5pt] table [y=FANG] {\MnistMedian};
\addplot [color=orange, mark=ball, mark size=0.5pt] table [y=LIE] {\MnistMedian};
\addplot [color=purple, mark=ball, mark size=0.5pt] table [y=minsum] {\MnistMedian};
\addplot [color=brown, mark=ball, mark size=0.5pt] table [y=mpaf] {\MnistMedian};
\addplot [color=teal, mark=otimes, mark size=0.5pt] table [y=PoisonedFL]{\MnistMedian};
\addplot [color=red, mark=square*, mark size=0.5pt] table [y=XFED_sign] {\MnistMedian};
\addplot [color=blue, mark=ball, mark size=0.5pt] table [y=XFED_unit_vec] {\MnistMedian};

\nextgroupplot[title={Purchase, FedAvg,\\Cross-silo},
  title style={font=\small,align=center}, 
  yticklabels=\empty, ylabel={}
]
\addplot [color=green, mark=ball, mark size=0.5pt] table [y=VIRAT_min_max] {\PurchaseMean};
\addplot [color=black, mark=ball, mark size=0.5pt] table [y=FANG] {\PurchaseMean};
\addplot [color=orange, mark=ball, mark size=0.5pt] table [y=LIE] {\PurchaseMean};
\addplot [color=purple, mark=ball, mark size=0.5pt] table [y=minsum] {\PurchaseMean};
\addplot [color=brown, mark=ball, mark size=0.5pt] table [y=mpaf] {\PurchaseMean};
\addplot [color=teal, mark=otimes, mark size=0.5pt] table [y=PoisonedFL]{\PurchaseMean};
\addplot [color=red, mark=square*, mark size=0.5pt] table [y=XFED_sign] {\PurchaseMean};
\addplot [color=blue, mark=ball, mark size=0.5pt] table [y=XFED_unit_vec] {\PurchaseMean};

\nextgroupplot[title={Purchase, Median,\\Cross-silo},
  title style={font=\small,align=center}, ylabel={\scriptsize Attack Impact},
  trim axis left,
  ]
\addplot [color=green, mark=ball, mark size=0.5pt] table [y=VIRAT_min_max] {\PurchaseMedian};
\addplot [color=black, mark=ball, mark size=0.5pt] table [y=FANG] {\PurchaseMedian};
\addplot [color=orange, mark=ball, mark size=0.5pt] table [y=LIE] {\PurchaseMedian};
\addplot [color=purple, mark=ball, mark size=0.5pt] table [y=minsum] {\PurchaseMedian};
\addplot [color=brown, mark=ball, mark size=0.5pt] table [y=mpaf] {\PurchaseMedian};
\addplot [color=teal, mark=otimes, mark size=0.5pt] table [y=PoisonedFL]{\PurchaseMedian};
\addplot [color=red, mark=square*, mark size=0.5pt] table [y=XFED_sign] {\PurchaseMedian};
\addplot [color=blue, mark=ball, mark size=0.5pt] table [y=XFED_unit_vec] {\PurchaseMedian};

\nextgroupplot[title={Purchase, FedAvg,\\Cross-device},
  title style={font=\small,align=center},
  yticklabels=\empty, ylabel={}
]
\addplot [color=red, mark=square*, mark size=0.5pt] table [y=XFED_sign] {\PurchaseCrossDeviceMean};
\addplot [color=blue, mark=ball, mark size=0.5pt] table [y=XFED_unit_vec] {\PurchaseCrossDeviceMean};

\nextgroupplot[title={Purchase, Median,\\Cross-device},
  title style={font=\small,align=center},
  yticklabels=\empty, ylabel={}
]
\addplot [color=red, mark=square*, mark size=0.5pt] table [y=XFED_sign] {\PurchaseCrossDeviceMedian};
\addplot [color=blue, mark=ball, mark size=0.5pt] table [y=XFED_unit_vec] {\PurchaseCrossDeviceMedian};

\end{groupplot}
\end{tikzpicture}

    \par\vspace{2mm}\centering
      \begin{tikzpicture}
        \begin{axis}[
            xmin=0, xmax=45,
            ymin=0, ymax=1,
            legend columns=4,
            axis lines=none,
            width=0.35\textwidth,
            height=3cm,
            clip=false,
            legend style={font=\scriptsize}, 
        ]
          \addlegendimage{color=blue, mark=ball, mark size=0.5pt}
            \addlegendentry{ \( X_{uv} \)}
          \addlegendimage{color=red, mark=square*, mark size=0.5pt}
            \addlegendentry{ \( X_{sgn} \)}
          \addlegendimage{color=green, mark=ball, mark size=0.5pt}
            \addlegendentry{\( Min-Max \)}
          \addlegendimage{color=black, mark=ball, mark size=0.5pt}
            \addlegendentry{\( Fang \)}
        \addlegendimage{color=orange, mark=ball, mark size=0.5pt}
            \addlegendentry{\( LIE\)}
        \addlegendimage{color=purple, mark=ball, mark size=0.5pt}
            \addlegendentry{\( Min-Sum\)}
             \addlegendimage{color=brown, mark=ball, mark size=0.5pt}
            \addlegendentry{\( MPAF\)}
             \addlegendimage{color=teal, mark=otimes, mark size=0.5pt}
            \addlegendentry{\( PoisonedFL\)}
        \end{axis}
      \end{tikzpicture}%

\caption{Attack Impact \( I_\theta \) on global model with increasing \% of malicious clients for different attacks and aggregations (FedAvg and Median) in different FL settings for different datasets (MNIST \& Purchase).}
\label{impact-Purchase-mnist-meam-median}
\end{figure}

\paragraph{Cross-Silo Setting} Figures \ref{impact-Purchase} and \ref{impact-Purchase-mnist-meam-median} illustrate the impact of model poisoning attacks on the learnt global model as the percentage of malicious clients varies from 1\% to 30\% for the Purchase and MNIST (FedAvg and Median) datasets. We observe that none of the state-of-the-art aggregations or defenses, except for FLTrust, can effectively mitigate our proposed \(X_{uv}\) attack across all percentages of malicious clients. Notably, the effectiveness of our \(X_{uv}\) attack increases as the fraction of malicious clients rises. Our attacks, along with other existing attacks, perform exceptionally well against FedAvg, as it is not robust against Byzantine faults and does not filter out malicious updates. However, FLTrust successfully mitigates the \(X_{uv}\) attack and other attacks by utilizing a small trusted root dataset on the server. On the other hand, FoolsGold is very unreliable; it shows poor performance with few attackers, but its effectiveness improves with a greater number of malicious clients. Still, its performance declines again at higher ratios of attackers.


\pgfplotstableread[col sep=comma, header=true]{Purchase-crossdevice-Krum.csv}\PurchaseCrossDeviceKrum

\pgfplotstableread[col sep=comma, header=true]{Purchase-crossdevice-TRMEAN.csv}\PurchaseCrossDeviceTrmean

\pgfplotstableread[col sep=comma, header=true]{Purchase-crossdevice-CC.csv}\PurchaseCrossDeviceCC

\pgfplotstableread[col sep=comma, header=true]{Purchase-crossdevice-SignGuard.csv}\PurchaseCrossDeviceSign

\pgfplotstableread[col sep=comma, header=true]{Purchase-crossdevice-FoolsGold.csv}\PurchaseCrossDeviceGold

\pgfplotstableread[col sep=comma, header=true]{Purchase-crossdevice-FreqFred.csv}\PurchaseCrossDeviceFreq

\pgfplotstableread[col sep=comma, header=true]{Purchase-crossdevice-FLAME.csv}\PurchaseCrossDeviceFLAME

\pgfplotstableread[col sep=comma, header=true]{Purchase-crossdevice-FLTrust.csv}\PurchaseCrossDeviceFLTrust

\begin{figure*}[tb]
\centering

\setlength\panelsep{2mm}

\begin{tikzpicture}
\begin{groupplot}[
  group style={group size=8 by 1, horizontal sep=\panelsep, vertical sep=0pt},
  width=\dimexpr(\textwidth - 9\panelsep)/8\relax,
  height=2.1cm,                 
  xtick={0,5,10,15,20,25,30},
  xticklabels={0,5,10,15,20,25,30},
  ytick style={draw=none},
  ymajorgrids=false,
  title style={yshift=-1mm},
  xmin=0,xmax=30,
  ymin=0,ymax=75, ytick={0,10,...,90}
]
\hspace{-5mm}
\nextgroupplot[
  title={\small M-Krum},
  ylabel={\scriptsize Attack Impact},
  trim axis left,
]
\addplot [color=red, mark=square*, mark size=0.5pt] table [y=XFED_sign] {\PurchaseCrossDeviceKrum};
\addplot [color=blue, mark=ball, mark size=0.5pt] table [y=XFED_unit_vec] {\PurchaseCrossDeviceKrum};

\nextgroupplot[title={\small Tr-Mean}, yticklabels=\empty, ylabel={}]
\addplot [color=red, mark=square*, mark size=0.5pt] table [y=XFED_sign] {\PurchaseCrossDeviceTrmean};
\addplot [color=blue, mark=ball, mark size=0.5pt] table [y=XFED_unit_vec] {\PurchaseCrossDeviceTrmean};

\nextgroupplot[title={\small CC}, yticklabels=\empty, ylabel={}]
\addplot [color=red, mark=square*, mark size=0.5pt] table [y=XFED_sign] {\PurchaseCrossDeviceCC};
\addplot [color=blue, mark=ball, mark size=0.5pt] table [y=XFED_unit_vec] {\PurchaseCrossDeviceCC};

\nextgroupplot[title={\small SgnG}, yticklabels=\empty, ylabel={}]
\addplot [color=red, mark=square*, mark size=0.5pt] table [y=XFED_sign] {\PurchaseCrossDeviceSign};
\addplot [color=blue, mark=ball, mark size=0.5pt] table [y=XFED_unit_vec] {\PurchaseCrossDeviceSign};

\nextgroupplot[title={\small FoolsGold}, yticklabels=\empty, ylabel={}]
\addplot [color=blue, mark=ball, mark size=0.5pt] table [y=XFED_unit_vec] {\PurchaseCrossDeviceGold};
\addplot [color=red, mark=square*, mark size=0.5pt] table [y=XFED_sign] {\PurchaseCrossDeviceGold};

\nextgroupplot[title={\small FLAME}, yticklabels=\empty, ylabel={}]
\addplot [color=red, mark=square*, mark size=0.5pt] table [y=XFED_sign] {\PurchaseCrossDeviceFLAME};
\addplot [color=blue, mark=ball, mark size=0.5pt] table [y=XFED_unit_vec] {\PurchaseCrossDeviceFLAME};

\nextgroupplot[title={\small FreqFed}, yticklabels=\empty, ylabel={}]
\addplot [color=red, mark=square*, mark size=0.5pt] table [y=XFED_sign] {\PurchaseCrossDeviceFreq};
\addplot [color=blue, mark=ball, mark size=0.5pt] table [y=XFED_unit_vec] {\PurchaseCrossDeviceFreq};

\nextgroupplot[title={\small FLTrust}, yticklabels=\empty, ylabel={}, trim axis right]
\addplot [color=red, mark=square*, mark size=0.5pt] table [y=XFED_sign] {\PurchaseCrossDeviceFLTrust};
\addplot [color=blue, mark=ball, mark size=0.5pt] table [y=XFED_unit_vec] {\PurchaseCrossDeviceFLTrust};

\end{groupplot}
\end{tikzpicture}

    \par\vspace{2mm}\centering
      \begin{tikzpicture}
        \begin{axis}[
            xmin=0, xmax=45,
            ymin=0, ymax=1,
            legend columns=8,
            axis lines=none,
            width=0.35\textwidth,
            height=3cm,
            clip=false,
            legend style={font=\scriptsize}, 
        ]
          \addlegendimage{color=blue, mark=ball, mark size=0.5pt}
            \addlegendentry{ \( X_{uv} \)}
          \addlegendimage{color=red, mark=square*, mark size=0.5pt}
            \addlegendentry{ \( X_{sgn} \)}
          \end{axis}
      \end{tikzpicture}%

\caption{Attack Impact \( I_\theta \) on global models with increasing \% of malicious clients for different aggregations, defenses, and attacks in a cross-device setting (Purchase dataset).}
\label{impact-Purchase-crossdevice}
\end{figure*}

\paragraph{Cross-Device Setting} Figures \ref{impact-Purchase-mnist-meam-median} and \ref{impact-Purchase-crossdevice} illustrate the impact of model poisoning attacks on the learned global model as the percentage of malicious clients varies from 1\% to 30\% in the cross-device setting for the Purchase dataset. To simulate this scenario, we consider 10,000 clients, with 1\% randomly selected in each round of federated learning (FL) training. We observe that the effectiveness of our \(X_{uv}\) attack increases as the fraction of malicious clients rises. Our attacks perform very well against FedAvg, as it is not Byzantine-robust and does not filter out malicious updates. However, \(X_{uv}\) is also very effective against other aggregations and defenses too. For instance, with the SignGuard aggregation rule, when the percentage of attackers is 5\%, the attack impact is 22.47; when the percentage of malicious clients increases to 20\%, the attack impact rises to 73.22. This indicates that the global model becomes nearly unusable as the percentage of malicious clients increases. Other aggregations and defenses, except for FLTrust, also show a similar trend.

\subsection{Impact of the degree of non-IID.}\label{noniid-appendix}

\paragraph{Non-IID Simulation} To simulate non-IID data distribution, we introduce a probabilistic model based on a parameter \( p \), which we refer to as the ``degree of non-IID''. We followed the process described in~\cite{fang2020local} to model this setting. At first, we evenly partition the total number of client devices into \( L \) groups, where \( L \) is the number of classes (e.g., \( L=10 \) for MNIST and \( L=100 \) for Purchase) in the classification problem. We then assign training instances with label \( l \) (where \( l \in \{1, 2, \ldots, L\} \)) to the \( l \)-th group with probability \( p \). A higher degree of \( p \) indicates training instances of class \( l \) are more concentrated in the \( l \)-th group, meaning datasets are not identically distributed. When \( p \) is smaller, the data distribution becomes more uniform across groups, approximating an IID scenario. 

\input{noniid-MNIST}
\pgfplotstableread[col sep=comma, header=true]{MNIST-noniid-Mean.csv}\MNISTMeanNoniid

\pgfplotstableread[col sep=comma, header=true]{MNIST-noniid-MEDIAN.csv}\MNISTMedianNoniid

\pgfplotstableread[col sep=comma, header=true]{Purchase-noniid-Mean.csv}\PurchaseMeanNoniid

\pgfplotstableread[col sep=comma, header=true]{Purchase-noniid-MEDIAN.csv}\PurchaseMedianNoniid

\begin{figure}[tb]
\centering

\setlength\panelsep{2mm} 

\begin{tikzpicture}
\begin{groupplot}[
  group style={
    group size=4 by 1,
    horizontal sep=6pt,
  },
  width=\dimexpr(\linewidth - 7\panelsep)/4\relax,
  height=2.1cm,    
  xtick={.1, .3, .5, .7, .9},
  xticklabels={.1, .3, .5, .7, .9},
  ytick style={draw=none},
  ymajorgrids=false,
  title style={font=\small, align=center},
  xmin=0.1,xmax=0.9,
  ymin=0,ymax=100, ytick={0,20,40,60,80,100}
]
 \hspace{-2mm}
\nextgroupplot[
  title={\small MNIST,\\Fed-Avg},
  ylabel={\scriptsize Attack Impact},
  trim axis left,
]
\addplot [color=magenta, mark=ball, mark size=0.5pt] table [y=NoAttack] {\MNISTMeanNoniid};
\addplot [color=green, mark=ball, mark size=0.5pt] table [y=VIRAT_min_max] {\MNISTMeanNoniid};
\addplot [color=black, mark=ball, mark size=0.5pt] table [y=FANG] {\MNISTMeanNoniid};
\addplot [color=orange, mark=ball, mark size=0.5pt] table [y=LIE] {\MNISTMeanNoniid};
\addplot [color=purple, mark=ball, mark size=0.5pt] table [y=minsum] {\MNISTMeanNoniid};
\addplot [color=brown, mark=ball, mark size=0.5pt] table [y=mpaf] {\MNISTMeanNoniid};
\addplot [color=teal, mark=otimes, mark size=0.5pt] table [y=PoisonedFL]{\MNISTMeanNoniid};
\addplot [color=red, mark=square*, mark size=0.5pt] table [y=XFED_sign] {\MNISTMeanNoniid};
\addplot [color=blue, mark=ball, mark size=0.5pt] table [y=XFED_unit_vec] {\MNISTMeanNoniid};

\nextgroupplot[title={\small MNIST,\\Median}, yticklabels=\empty, ylabel={}]
\addplot [color=magenta, mark=ball, mark size=0.5pt] table [y=NoAttack] {\MNISTMedianNoniid};
\addplot [color=green, mark=ball, mark size=0.5pt] table [y=VIRAT_min_max] {\MNISTMedianNoniid};
\addplot [color=black, mark=ball, mark size=0.5pt] table [y=FANG] {\MNISTMedianNoniid};
\addplot [color=orange, mark=ball, mark size=0.5pt] table [y=LIE] {\MNISTMedianNoniid};
\addplot [color=purple, mark=ball, mark size=0.5pt] table [y=minsum] {\MNISTMedianNoniid};
\addplot [color=brown, mark=ball, mark size=0.5pt] table [y=mpaf] {\MNISTMedianNoniid};
\addplot [color=teal, mark=otimes, mark size=0.5pt] table [y=PoisonedFL]{\MNISTMedianNoniid};
\addplot [color=red, mark=square*, mark size=0.5pt] table [y=XFED_sign] {\MNISTMedianNoniid};
\addplot [color=blue, mark=ball, mark size=0.5pt] table [y=XFED_unit_vec] {\MNISTMedianNoniid};

\nextgroupplot[title={\small Purchase,\\Fed-Avg}, yticklabels=\empty, ylabel={}
  ]
\addplot [color=magenta, mark=ball, mark size=0.5pt] table [y=NoAttack] {\PurchaseMeanNoniid};
\addplot [color=green, mark=ball, mark size=0.5pt] table [y=VIRAT_min_max] {\PurchaseMeanNoniid};
\addplot [color=black, mark=ball, mark size=0.5pt] table [y=FANG] {\PurchaseMeanNoniid};
\addplot [color=orange, mark=ball, mark size=0.5pt] table [y=LIE] {\PurchaseMeanNoniid};
\addplot [color=purple, mark=ball, mark size=0.5pt] table [y=minsum] {\PurchaseMeanNoniid};
\addplot [color=brown, mark=ball, mark size=0.5pt] table [y=mpaf] {\PurchaseMeanNoniid};
\addplot [color=teal, mark=otimes, mark size=0.5pt] table [y=PoisonedFL]{\PurchaseMeanNoniid};
\addplot [color=red, mark=square*, mark size=0.5pt] table [y=XFED_sign] {\PurchaseMeanNoniid};
\addplot [color=blue, mark=ball, mark size=0.5pt] table [y=XFED_unit_vec] {\PurchaseMeanNoniid};

\nextgroupplot[title={\small Purchase,\\Median}, yticklabels=\empty, ylabel={}]
\addplot [color=magenta, mark=ball, mark size=0.5pt] table [y=NoAttack] {\PurchaseMedianNoniid};
\addplot [color=green, mark=ball, mark size=0.5pt] table [y=VIRAT_min_max] {\PurchaseMedianNoniid};
\addplot [color=black, mark=ball, mark size=0.5pt] table [y=FANG] {\PurchaseMedianNoniid};
\addplot [color=orange, mark=ball, mark size=0.5pt] table [y=LIE] {\PurchaseMedianNoniid};
\addplot [color=purple, mark=ball, mark size=0.5pt] table [y=minsum] {\PurchaseMedianNoniid};
\addplot [color=brown, mark=ball, mark size=0.5pt] table [y=mpaf] {\PurchaseMedianNoniid};
\addplot [color=teal, mark=otimes, mark size=0.5pt] table [y=PoisonedFL]{\PurchaseMedianNoniid};
\addplot [color=red, mark=square*, mark size=0.5pt] table [y=XFED_sign] {\PurchaseMedianNoniid};
\addplot [color=blue, mark=ball, mark size=0.5pt] table [y=XFED_unit_vec] {\PurchaseMedianNoniid};

\end{groupplot}
\end{tikzpicture}

    \par\vspace{2mm}\centering
      \begin{tikzpicture}
        \begin{axis}[
            xmin=0, xmax=45,
            ymin=0, ymax=1,
            legend columns=4,
            axis lines=none,
            width=0.28\textwidth,
            height=3cm,
            clip=false,
            legend style={font=\scriptsize}, 
        ]
                  \addlegendimage{color=magenta, mark=ball, mark size=0.5pt}
            \addlegendentry{ \(  A_\theta \)}
          \addlegendimage{color=blue, mark=ball, mark size=0.5pt}
            \addlegendentry{ \( X_{uv} \)}
          \addlegendimage{color=red, mark=square*, mark size=0.5pt}
            \addlegendentry{ \( X_{sgn} \)}
          \addlegendimage{color=green, mark=ball, mark size=0.5pt}
            \addlegendentry{\( Min-Max \)}
          \addlegendimage{color=black, mark=ball, mark size=0.5pt}
            \addlegendentry{\( Fang \)}
        \addlegendimage{color=orange, mark=ball, mark size=0.5pt}
            \addlegendentry{\( LIE\)}
        \addlegendimage{color=purple, mark=ball, mark size=0.5pt}
            \addlegendentry{\( Min-Sum\)}
             \addlegendimage{color=brown, mark=ball, mark size=0.5pt}
            \addlegendentry{\( MPAF\)}
             \addlegendimage{color=teal, mark=otimes, mark size=0.5pt}
            \addlegendentry{\( PoisonedFL\)}
        \end{axis}
      \end{tikzpicture}%

\caption{Attack Impact \( I_\theta \) on global model as a function of the degree of non-IID for different attacks and aggregations (FedAvg and Median) for different datasets (MNIST \& Purchase).}
  \label{fig:mean-median-purchase-mnist-noniid}
\end{figure}
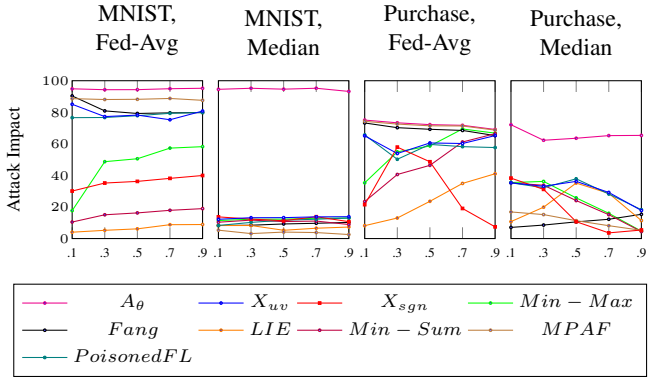

\paragraph{Impact of the degree of non-IID} Fig.~\ref{fig:non-iid-MNIST} and Fig.~\ref{fig:mean-median-purchase-mnist-noniid} show the impact of different attacks on the global model as \(p\) increases from 0.1 to 0.9 on the MNIST and Purchase (FedAvg \& Median) datasets. We observe that \(X_{uv}\) outperforms all state-of-the-art attacks by a significant margin for Clipped Clustering, SignGuard, FoolsGold, FLAME, and FreqFed. As \(p\) increases and the data become more unevenly distributed across clients, the performance of \(X_{uv}\) does not degrade; the attack impact stays roughly the same for all levels of non-IID data. For M-Krum and Trimmed-Mean, \(X_{uv}\) also works very well; however, for Trimmed-Mean, the Fang attack sometimes has slightly higher impact, which is reasonable because Fang is specifically designed for that aggregation rule. Overall, even for higher values of \(p\), \(X_{uv}\) remains robust.

 \subsection{\textbf{Performance variations of model poisoning attacks across Neural Network architectures}} \label{diff-neural-network-appendix}

We explore different Neural Network architectures for the Fashion-MNIST dataset and compare the performance of \(X_{sgn}\) and \(X_{uv}\) against state-of-the-art model poisoning attacks. The results are presented in Table~\ref{tab:main-results} (c) for a 3-layer DNN, while Table~\ref{tab:main-results} (e) displays the results for the AlexNet architecture. Our results show that \(X_{uv}\) attack is highly effective across various neural network architectures. The effectiveness of our attack is not dependent on the specific model architecture; instead, our attack exploits the vulnerabilities inherent in federated learning aggregation. Thus, using a deeper neural network does not safeguard an FL system, and a non-collusive attack like \(X_{uv}\) can reliably outperform existing state-of-the-art attacks, regardless of the model architecture used in the FL system.

\subsection{Influence of various attack parameters}
\label{attack-parameter-appendix}

\begin{table*}[!tb]
\centering
\caption{Attack impact \(I_\theta\) of the global model under the \(X_{uv}\) attack (20\% malicious clients) for different values of \(\Omega\) across various aggregation rules and defenses on the MNIST dataset in the cross-silo setting. In all experiments, we set \(\lambda = 4\).}
\label{tab:omega-table-appendix}
\resizebox{\textwidth}{!}{%
\begin{tabular}{@{}l||ccccccccccc@{}}
\toprule
\begin{tabular}[c]{@{}c@{}}Aggregation \\ Rule\end{tabular} & \multicolumn{1}{l}{\textbf{FedAvg}} & \multicolumn{1}{l}{\textbf{Median}} & \multicolumn{1}{l}{\textbf{Tr-Mean}} & \multicolumn{1}{l}{\textbf{Multi-Krum}} & \multicolumn{1}{l}{\textbf{Clipped-Clustering}} & \multicolumn{1}{l}{\textbf{SignGuard}} & \multicolumn{1}{l}{\textbf{FLTrust}} & \multicolumn{1}{l}{\textbf{FreqFed}} & \multicolumn{1}{l}{\textbf{FoolsGold}} & \multicolumn{1}{l}{\textbf{FLAME}} \\ \midrule
\multicolumn{1}{c||}{\( A_\theta \)}        & 95.02                             & 94.61                      & 94.69                            & 94.57                          & 94.08                                  & 95.31                         & 95.49                       & 94.32                       & 95.12                         & 93.33                     \\ \midrule
\multicolumn{1}{c||}{$\Omega = 4$}                                   & 83.71                             & 11.15                      & 14.99                            & 12.5                           & 78.08                                  & 79.16                         & 3.26                        & 84.5                        & 36.11                         & 76.17                     \\
\multicolumn{1}{c||}{$\Omega = 8$}                                   & 77.26                             & 13.41                      & 15.48                            & 12.56                          & 82.89                                  & 78.41                         & 7.46                        & 82.16                       & 34.5                          & 74.82                     \\
\multicolumn{1}{c||}{$\Omega = 12$}                                   & 78.47                             & 11.96                      & 13.49                            & 13.22                          & 81.28                                  & 84.39                         & 3.49                        & 76.5                        & 29.49                         & 76.88                     \\
\multicolumn{1}{c||}{$\Omega = 16$}                                    & 84.82                             & 11.45                      & 12.98                            & 11.22                          & 76.87                                  & 79.67                         & 4.19                        & 79.59                       & 34.14                         & 78.69                     \\
\multicolumn{1}{c||}{$\Omega = 20$}                                   & 76.45                             & 12.15                      & 15.72                            & 12.52                          & 79.08                                  & 80.16                         & 1.79                        & 76.91                       & 31.08                         & 70.83                     \\
\multicolumn{1}{c||}{$\Omega = 24$}                                    & 74.72                             & 11.64                      & 14.39                            & 12.89                          & 75.68                                  & 82.77                         & 4.21                        & 81.26                       & 28.75                         & 75.67                     \\
\multicolumn{1}{c||}{$\Omega = 28$}                                   & 77.16                             & 12.62                      & 16.83                            & 11.98                          & 74.07                                  & 78.75                         & 6.66                        & 83.55                       & 29.5                          & 79.53                     \\
\multicolumn{1}{c||}{$\Omega = 32$}                                    & 74.2                              & 11.23                      & 13.87                            & 12.96                          & 77.87                                  & 84.03                         & 2.63                        & 84.52                       & 36.59                         & 73.13                     \\
\multicolumn{1}{c||}{$\Omega = 36$}                                    & 75.03                             & 13.11                      & 14.7                             & 11.85                          & 80.08                                  & 81.16                         & 3.87                        & 77.54                       & 36.02                         & 77.84                     \\ \bottomrule
\end{tabular}%
}
\end{table*}

\pgfplotstableread[col sep=comma, header=true]{Omega-ExecutionTime.csv}\MnistOmegaTiming

\begin{figure}[htb]
    \centering
    
 
      \begin{tikzpicture}
        \begin{axis}[
            ylabel={\scriptsize{Attack Execution Time (s)}},
            xlabel={\scriptsize{Value of \(\Omega\)}},
            xmin=4, xmax=260,
            ymin=1600, ymax=2000,
            xtick={20, 60, 100, 140, 180, 220, 260},
            ytick={1600, 1700, 1800, 1900, 2000},
            legend pos=north west,
            width=0.55\linewidth
        ]
          \addplot [color=blue, mark=ball, mark size=1pt] table [y=Time] {\MnistOmegaTiming};

        \end{axis}
      \end{tikzpicture}%

    \caption{Impact of the history window size \(\Omega\) on the execution time of the \(X_{uv}\) attack (20\% malicious clients) on the MNIST dataset in the cross-silo setting. As \(\Omega\) increases, the runtime grows noticeably.}
    \label{fig:omega-timing-graph}
\end{figure}
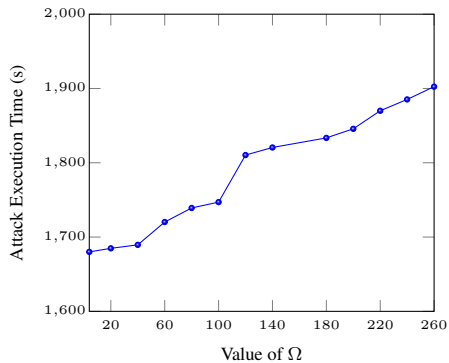

\paragraph{\textbf{Impact of the size of $\mathcal{H}^{(t)}$}} Recall that for the global model \( \theta^{(t)} \) at round \( t \), we define its delta from the previous round as
\[
\Delta\theta^{(t)} = \theta^{(t)} - \theta^{(t-1)},
\]
and maintain a list of all deltas
\[
\mathcal{H}^{(t)} = \{\, \Delta\theta^{(1)}, \ldots, \Delta\theta^{(t)} \,\}.
\]

Now in FL, the number of training rounds can go on for a long time. With each round $t$, $\mathcal{H}^{(t)}$ size increase. For a client device, it may not be practical to store so many global deltas. Especially if the attacker is a silent malware type hiding on a mobile device. In practice, we suggest that an attacker store the last $\Omega$ global deltas. Therefore, the modified global delta list $\mathcal{H}^{(t)}$ be as,

\[
\mathcal{H}^{(t)}=\{\Delta\theta^{(t-\Omega+1)},\ldots,\Delta\theta^{(t)}\}
\]

This might seem like a loss of information, but for practical implementation, it would be needed. We study how \( \Omega \) affects the impact of our proposed \(X_{uv}\) attack. Table~\ref{tab:omega-table-appendix} reports the attack impact \(I_\theta\) of the global model for the \(X_{uv}\) attack (20\% malicious clients) for different values of \(\Omega\) across various aggregation rules and defenses on the MNIST dataset. We vary \(\Omega\) from 4 to 36 for all aggregations and defenses, and in all experiments, we fix \(\lambda = 4\). From the table, we observe that the attack impact is largely insensitive to \(\Omega\). For example, for the FreqFed defense, the impact changes only from 84.50 to 77.54, and for FoolsGold, it varies slightly from 36.11 to 36.02 as \(\Omega\) ranges from 4 to 36. These small variations indicate that the number of stored deltas \(\Omega\) does not significantly affect the effectiveness of the attack. In contrast, \(\Omega\) clearly affects the attack's execution time. As shown in Fig.~\ref{fig:omega-timing-graph}, increasing \(\Omega\) increases the runtime; when \(\Omega = 4\), the execution time is 1680.02 seconds, whereas for \(\Omega = 180\) it increases to 1833.48 seconds. Therefore, a small value such as \(\Omega = 4\) is practically preferable, as it achieves almost the same attack impact while reducing computational cost, and we adopt \(\Omega = 8\) as our default choice.

\pgfplotstableread[col sep=comma, header=true]{lambda-impact.csv}\MnistLambda

\begin{figure}[htb]
    \centering
    
      \begin{tikzpicture}[baseline=(current bounding box.north)]
        \begin{axis}[
            ylabel={\scriptsize{Attack Impact}},
            xlabel={\scriptsize{Value of \(\lambda\)}},
            xmin=2, xmax=10,
            ymin=0, ymax=90,
            xtick={2, 4, 6, 8, 10},
            ytick={0, 10, 20, 30, 40, 50, 60, 70, 80, 90},
            legend pos=north west,
            width=0.55\linewidth
        ]
      
\addplot [color=magenta, mark=ball, mark size=0.5pt] table [y=FedAvg] {\MnistLambda};
\addplot [color=green, mark=ball, mark size=0.5pt] table [y=Median] {\MnistLambda};
\addplot [color=black, mark=ball, mark size=0.5pt] table [y=TrimmedMean] {\MnistLambda};
\addplot [color=orange, mark=ball, mark size=0.5pt] table [y=Multi-Krum] {\MnistLambda};
\addplot [color=purple, mark=ball, mark size=0.5pt] table [y=CC] {\MnistLambda};
\addplot [color=brown, mark=ball, mark size=0.5pt] table [y=SignGuard] {\MnistLambda};
\addplot [color=teal, mark=otimes, mark size=0.5pt] table [y=FLTrust]{\MnistLambda};
\addplot [color=red, mark=square*, mark size=0.5pt] table [y=FreqFed] {\MnistLambda};
\addplot [color=blue, mark=ball, mark size=0.5pt] table [y=FoolsGold] {\MnistLambda};
\addplot [color=yellow, mark=ball, mark size=0.5pt] table [y=FLAME] {\MnistLambda};

        \end{axis}
      \end{tikzpicture}
      \raisebox{-3mm}{%
          \begin{tikzpicture}[baseline=(current bounding box.north)]
          \begin{axis}[
              xmin=0, xmax=45,
              ymin=0, ymax=1,
              legend columns=1,
              axis lines=none,
              width=0.2\textwidth,
              height=3cm,
              clip=false,
              legend style={font=\scriptsize},
          ]
               \addlegendimage {color=magenta, mark=ball, mark size=0.5pt} \addlegendentry{FedAvg}
                \addlegendimage {color=green, mark=ball, mark size=0.5pt} \addlegendentry{Median}
                \addlegendimage {color=black, mark=ball, mark size=0.5pt} \addlegendentry{TrimmedMean}
                \addlegendimage {color=orange, mark=ball, mark size=0.5pt} \addlegendentry{Multi-Krum}
                \addlegendimage {color=purple, mark=ball, mark size=0.5pt} \addlegendentry{CC}
                \addlegendimage {color=brown, mark=ball, mark size=0.5pt} \addlegendentry{SignGuard}
                \addlegendimage {color=teal, mark=otimes, mark size=0.5pt} \addlegendentry{FLTrust}
                \addlegendimage {color=red, mark=square*, mark size=0.5pt} \addlegendentry{FreqFed}
                \addlegendimage {color=blue, mark=ball, mark size=0.5pt} \addlegendentry{FoolsGold}
                \addlegendimage {color=yellow, mark=ball, mark size=0.5pt} \addlegendentry{FLAME}
            \end{axis}
          \end{tikzpicture}%
      }

    \caption{Effect of \(\lambda\) on the attack impact \(I_\theta\) of the \(X_{uv}\) attack (20\% malicious clients) on the MNIST dataset in cross-silo setting.  In all experiments, we set \(\Omega = 8\).}
    \label{fig:lambda-graph}
\end{figure}
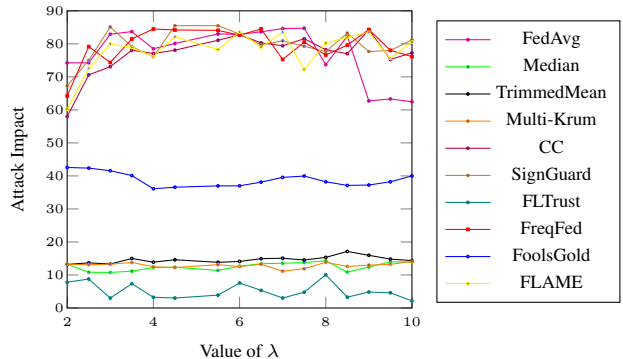

\paragraph{Impact of $\lambda$} As discussed in Appendix~\ref{lambda-limit}, the recommended range for $\lambda$, derived from outlier detection guidelines in the literature and common practice, is $[2, 10]$. Fig.~\ref{fig:lambda-graph} shows the impact of the $X_{uv}$ attack across this range, and we observe that, within $[2, 10]$, the attack impact remains highly stable. The attack impact of \(X_{uv}\) for FedAvg, Clipped-Clustering, SignGuard, FreqFed, and FLAME decreases slightly around \(\lambda = 2\) and \(\lambda = 10\), but remains largely stable between \(3\) and \(7\), with no substantial loss in effectiveness. For the other aggregations and defenses, choosing \(\lambda \in [2, 10]\) does not affect the attack impact. However, as shown in Table~\ref{tab:lambda-extream}, taking extreme values of \(\lambda\), such as \(0\) or \(100\), can substantially reduce the attack impact.

\begin{table}[h]
    \centering
    \small 
    \caption{Effect of extreme values of \(\lambda\) on the attack impact \(I_\theta\) of the \(X_{uv}\) attack (20\% malicious clients) on the MNIST dataset.}
    \label{tab:lambda-extream}
    \begin{tabular}{@{}l||lccc}
        \toprule
        Aggregation Rules \& Defenses & $\lambda = 0$ & $\lambda = 4$ & $\lambda = 100$ \\
        \midrule
        \multicolumn{1}{c||}{FedAvg}             & 24.64 & 81.43 & 10    \\
        \multicolumn{1}{c||}{Median}             & 9.35  & 12.21 & 11    \\
        \multicolumn{1}{c||}{Tr-Mean}            & 12.15 & 15.19 & 12.6  \\
        \multicolumn{1}{c||}{M-Krum}             & 4.11  & 12.5  & 1.23  \\
        \multicolumn{1}{c||}{Clipped-Clustering} & 24.21 & 80.1  & 2.45  \\
        \multicolumn{1}{c||}{SignGuard}          & 23.53 & 81.18 & 4.51  \\
        \multicolumn{1}{c||}{FLTrust}            & 5.19  & 7.35  & 3.19  \\
        \multicolumn{1}{c||}{FreqFed}            & 22.65 & 80.87 & 9.16  \\
        \multicolumn{1}{c||}{FoolsGold}          & 2.97  & 36.11 & 2.86  \\
        \multicolumn{1}{c||}{FLAME}              & 34.21 & 79.24 & 2.39  \\
        \bottomrule
    \end{tabular}
\end{table}

\paragraph{Implementation suggestion for XFED} Based on our experiments, we recommend implementing XFED using the \(X_{uv}\) attack with \(\Omega = 8\) and \(\lambda = 4\). This configuration provides a practical rule of thumb for deploying the attack. The authors of~\cite{shejwalkar2021manipulating} propose a method for selecting effective perturbation vectors via small-scale local simulations, which can be adopted as a general guideline when designing or evaluating attack strategies. In practice, it is advisable to follow their methodology to determine the most effective combination of perturbation vector and hyperparameters before launching an attack.

\subsection{FLTrust Root Dataset Bias}
\label{fltrust-root-bias}

In FLTrust~\cite{cao2021fltrust}, a trusted server maintains a small, clean \emph{root dataset} that is used to compute a reference gradient in each communication round. Client updates are then compared against this trusted gradient to assess their similarity. However, the performance of FLTrust is highly sensitive to the class distribution of the root dataset, characterized by the \emph{bias probability} in the original paper. This bias probability is defined as the fraction of examples in the root dataset that are sampled from a single class; a higher value therefore models a root data distribution that deviates more strongly from the overall training distribution. Empirical results show that FLTrust remains accurate and robust when the bias probability is not too large. 

\begin{table}[h]
\centering
\caption{Effect of root-data bias probability on FLTrust performance for the Purchase dataset. The table reports test accuracy under no attack and the attack impact \(I_\theta\) (20\% malicious clients) for $X_{uv}$ attack with different bias probabilities, showing that higher bias in the root dataset degrades clean accuracy and increases vulnerability to attacks.}
\label{tab:bias-table}
\resizebox{\columnwidth}{!}{%
\begin{tabular}{@{}l||ccc}
\toprule
Bias Probability & No Attack Accuracy & Attack Impact (20\% malicious clients) \\ 
\midrule
\multicolumn{1}{c||}{0.1} & 73.91 & 8.83  \\
\multicolumn{1}{c||}{0.3} & 71.52 & 9.95  \\
\multicolumn{1}{c||}{0.5} & 68.50 & 10.25 \\
\multicolumn{1}{c||}{0.7} & 61.90 & 14.82 \\
\multicolumn{1}{c||}{0.9} & 60.22 & 17.11 \\
\bottomrule
\end{tabular}
}
\end{table}

Table~\ref{tab:bias-table} summarizes the effect of the root-data bias probability on FLTrust for the Purchase dataset. As the bias probability increases, the clean (``No Attack'') test accuracy consistently degrades: at 0.1, it is 73.91\%; at 0.3, 71.52\%; at 0.5, 68.50\%; at 0.7, 61.90\%; and at 0.9, 60.22\%. At the same time, the attack impact \(I_\theta\) becomes larger as the bias increases. For example, when the bias probability is 0.1, the attack impact is 8.83, but it rises to 14.82 at 0.7 and 17.11 at 0.9. Therefore, a more biased root dataset lowers the baseline performance and makes FLTrust more vulnerable to model poisoning attacks. Consequently, in all our FLTrust experiments, we set the bias probability to 0.1, a setting in which FLTrust can effectively mitigate all state-of-the-art model poisoning attacks considered in this work.

\end{document}